\documentclass{article}
\usepackage[english]{babel}
\usepackage{amsmath,amssymb}
\usepackage{jheppub} 

\newcommand{\tmop}[1]{\ensuremath{\operatorname{#1}}}

\begin{document}

{\flushright Imperial-TP-2025-CH-2\\[1cm]}

\vspace{2cm}

\begin{center}
{\bf \Large Duality Symmetries}\\[7mm]
  Chris Hull \\[4mm]
{\sl  The Blackett Laboratory, Imperial College London, \\ Prince Consort Road, \\ London, SW7 2AZ, UK
\\[1mm]
 Email: {\tt c.hull@imperial.ac.uk}} \\[17mm]
 \end{center}
\textbf{Abstract:} Duality symmetries in supergravity theories, together with their implications for string theory, are reviewed.

\vspace{8cm}

\noindent
*Invited contribution to the book {\it Half a Century of Supergravity}, edited by Anna Ceresole and Gianguido Dall’Agata.

\newpage

\section{Introduction}

In 1978, Cremmer and Julia constructed $\mathcal{N}= 8$ supergravity in $d =
4$ dimensions and showed that the field equations are invariant under the
action of a global non-compact $E_7$ symmetry, with some group elements acting
via electromagnetic duality transformations \cite{Cremmer:1978ds,Cremmer:1979up}. This large exceptional
symmetry came as a great surprise. As the theory was constructed by toroidal
compactification from 11 dimensions, invariance under $G L (7, \mathbb{R})$
was to be expected, together with further abelian symmetries coming from
shifts of the 3-form gauge field, but the origin of the remaining symmetries,
which emerge after dualising come of the gauge fields, was mysterious and led
to these being termed \textit{hidden symmetries}. It was also one of the
first times that physicists had been led to consider a role for an exceptional
Lie group.

This was then extended to all supergravity theories with at least 16 local
supersymmetries in all dimensions $d < 11$, which all have a
rigid non-compact symmetry group $G.$ 
For 5-dimensional $\mathcal{N}= 8$ supergravity  $G=E_6$ \cite{Cremmer:1980gs}.
For the maximal supergravities
constructed by compactification of 11-dimensional supergravity on an
$n$-torus, the symmetry group is in the exceptional $E$ series with $G = E_n$
\cite{Julia:1980gr,Julia:1981wc}; see table 1.
It is a maximally non-compact form of $E_n$, sometimes
denoted as $E_{n (+ n)}$, where the difference between the number of
non-compact generators $N_n$ and the number of compact generators $N_c$ is
$N_n - N_c =  n$. Here
\[ E_5 = S p i n (5, 5), \quad E_4 = S L (5, \mathbb{R}), \quad E_3 = S L (3,
   \mathbb{R}) \times S L (2, \mathbb{R}), \quad E_2 = S L (2, \mathbb{R})
   \times \mathbb{R}^+, \quad E_1 =\mathbb{R}^+ \]

In each case, the scalar fields take values in a coset space $G / H$ where $H$
is the maximally compact subgroup of $G$, and $G$ acts on the scalar fields
via the natural action of $G$ on $G / H$. For example, for $G = E_7$, $H = S U
(8) /\mathbb{Z}_2$. In some cases $G$ is a symmetry of the action, while in
others it action includes electromagnetic duality transformations and so is
only a symmetry of the equations of motion. As the first examples included
electromagnetic dualities, the symmetries are often referred to as
\textit{duality symmetries}. Some of the symmetry in $d$-dimensions arise
from symmetries in 10 or 11 dimensions, but others mix fields arising from the
11-dimensional metric with fields arising from the 11-dimensional 3-form, and
so do not correspond to local symmetries of the 11-dimensional supergravity.

Each supergravity theory can be put in different dual forms by dualising
$p$-form gauge fields in $d$ dimensions to $\tilde{p}$-form gauge fields
$\tilde{p} = d - p - 2$ . The version of the theory with explicit $E_n$
symmetry arises from dualising so that each field has minimal degree, so that
a $\tilde{p}$-form field is dualised to a $p$-form field whenever $\tilde{p} >
p$. Other dual versions of the theory can have different duality groups, and
the duality group need not be a subgroup of $E_n$; explicit forms of the
dualities for other versions are given in \cite{Cremmer:1997ct}.
The duality group depends on the compactification manifold, e.g.\ compactifying 11-dimensional supergravity on $T^4$ gives $G=SL(5,\mathbb{R})$ while  on $K3$ gives $G=SO(3,19)$. For this reason, not all of the duality group can be understood in terms of the higher dimensional theory.


A solution of the gravity or supergravity equations in $D$ dimensions with $n$
commuting Killing vectors is also a solution of the dimensionally reduced
equations in $d = D - n$ dimensions. If the Killing vectors are all spacelike
this will be the reduction of the theory on $T^n$ or $\mathbb{R}^n$ 
(or $\mathbb{R}^r \times T^{n - r}$), 
while if one Killing vector is timelike and
the rest are spacelike, it will be a timelike reduction discussed in 
section \ref{timelike}, which also gives a duality $G$. The action of the duality symmetry $G$ of the equations of motion in
$d$ dimensions can then be used to generate new solutions. For example, for
gravity in 3+1 dimensions with one Killing vector, the $G=S L (2, \mathbb{R})$
Ehlers transformations discussed in section \ref{Ehlers} can be used to generate new solutions from old ones;
e.g.\ the mass parameter is transformed into the NUT charge, so that a
solution with mass can be transformed into a solution with both mass and NUT
charge \cite{Gibbons:1979xm}. This has been explored extensively 
in \cite{Breitenlohner:1987dg} and used to derive many properties of stationary
solutions.

Classical global symmetries of a theory including gravity are not expected to
survive in the quantum theory. In \cite{Hull:1994ys}, it was shown that for any supersymmetric
quantum theory whose low energy effective action is a supergravity with at
least 16 supersymmetries, the symmetry $G$ is broken to a discrete subgroup $G
(\mathbb{Z})$. In particular, for toroidal compactifications of superstring
theory or M-theory, $G (\mathbb{Z})$ is a discrete gauge symmetry, named
\textit{U--duality} in \cite{Hull:1994ys}. This duality symmetry, which combines
diffeomorphisms and T-dualities with non-perturbative symmetries, has proved
very important in understanding the non-perturbative structure of superstring
theory/M-theory. In particular, U-duality implies that non-perturbative
superstring theory is not just a theory of strings but also a theory of
branes, and as the branes are dual to strings, the branes should be regarded
as being on the same footing as the strings \cite{Hull:1994ys}.

There is an extensive literature on duality symmetries in supergravity -- for
a recent review, see \cite{Samtleben:2023ivs}, and for a history of the development of supergravity duality, see Bernard Julia's contribution to this volume. In this review, many of the technical details
that are readily available in the literature will not be repeated, but instead the focus will be on the key ideas.

\section{Duality Symmetry in Supergravity} \label{dualsugra}

In 10 dimensions there can be $(p, q)$ supersymmetry with $p$ right-handed
Majorana-Weyl supercharges and $q$ left-handed ones. There are two
supergravities with 32 supersymmetries, the IIA theory with (1,1)
supersymmetry that is obtained by circle reduction of 11-dimensional
supergravity and the IIB supergravity with (2,0) supersymmetry. The maximal
supergravity theories in $d = 11 - n \leqslant 9$ are unique and the same
theory is obtained by toroidal compactification of IIA or IIB supergravities
on $T^{n - 1}$ or of 11-dimensional supergravity on $T^n$.
The duality symmetries for these theories are listed in  table 1.

\begin{table}
\[ \begin{array}{|c|c|c|c|}
     \hline
     d & G & H & \tilde{H}\\
     \hline
     10 \tmop{IIB} & S L (2, \mathbb{R}) & S O (2) & S p i n (2)\\
     \hline
     10 \tmop{IIA} & \mathbb{R}^+ & 1 & 1\\
     \hline
     9 & S L (2, \mathbb{R}) \times \mathbb{R}^+ & S O (2) & S p i n (2)\\
     \hline
     8 & S L (3, \mathbb{R}) \times S L (2, \mathbb{R}) & S O (3) \times S O
     (2) & S p i n (3) \times S p i n (2)\\
     \hline
     7 & S L (5, \mathbb{R}) & S O (5) & S p i n (5) = U S p (4)\\
     \hline
     6 & S p i n (5, 5) & U S p (4) \times U S p (4) /\mathbb{Z}_2 & U S p (4)
     \times U S p (4)\\
     \hline
     5 & E_6 & U S p (8) /\mathbb{Z}_2 & U S p (8)\\
     \hline
     4 & E_7 & S U (8) /\mathbb{Z}_2 & S U (8)\\
     \hline
     3 & E_8 & S p i n (16) /\mathbb{Z}_2 & S p i n (16)\\
     \hline
   \end{array} \]
   \caption{Duality groups of maximal supergravities in $d$ dimensions.}
\end{table}

The scalars take values in the coset space $G / H$. Each theory can be formulated in a way that has a global $G$
symmetry that acts on the bosons (apart from the graviton) and not the
fermions, together with a local $\tilde{H}$ symmetry, where $\tilde{H}$ is the
double cover of the maximal compact subgroup $H$ of $G$. The group $\tilde{H}$
acts on the fermions and the scalars but not the other bosons.

The scalar fields can be represented by a $G$-valued scalar field
\[ \mathcal{V} (x) \in G \]
that transforms under a global $G$ transformation acting through $g \in G$ and
a local 
$H$ transformation acting through $h(x)\in H$:
\begin{equation}
  \mathcal{V} \rightarrow h (x)  \mathcal{V}g  \label{vtrans}
\end{equation}
There are then $d_G = \dim (G)$ scalar fields represented by $\mathcal{V} (x)$
and $d_H = \dim (H)$ gauge invariances that can be used to set $d_H$ of the
scalar fields to zero, leaving $d_G - d_H$ scalar fields parameterising $G / H$.
For example,
for $d = 4,$ $\mathcal{N}= 8$, the theory can be formulated in terms of 133
scalars taking values in $G = E_7$ and the 63-dimensional local symmetry group
$H = S U (8) /\mathbb{Z}_2$ can be used to set 63 scalars to zero, leaving 70
physical scalars. 

A local $\tilde{H}$ transformation $\tilde{h} (x) \in \tilde{H}$ acts on the scalars through $h (x) = \pi (\tilde{h} (x))$ where $\pi : \tilde{H} \rightarrow H$ is
the double cover map. The gauge choice setting  $d_H$ of the
scalar fields to zero
  fixes the local $H$ invariance, but reduces the $\tilde{H}$ invariance to
a $\mathbb{Z}_2$ symmetry $- 1^F$ that acts as $- 1$ on each fermion, $\psi
\rightarrow - \psi$. In this gauge, a $G$ transformation must be accompanied
by an $H$ transformation in order to maintain the gauge, and as a result all
fields except the graviton transform under $G$ in the gauge-fixed theory. The
$\tilde{H}$ invariance is sometimes referred to as R-symmetry.

The theories can be further compactified to give maximal supergravity theories
in $d < 3$ dimensions with duality groups $E_{11 - d}$ which are all infinite
dimensional. In $d = 2$ the duality group is $E_9$ which is affine $E_8$ (with
Lie algebra the $E_8 $ Kac-Moody algebra). The group $E_{10}$ and its
applications have been discussed in \cite{Damour:2002cu} while $E_{11}$ and its applications
have been discussed in \cite{West:2003fc}.

These theories can be truncated to theories with fewer supersymmetries. For
example, in $d = 4$ the pure supergravities with $\mathcal{N} \geqslant 4$
have the following duality groups listed in table 2. In each case, the R-symmetry group is $S U (\mathcal{N})$ or $U
(\mathcal{N})$.

\begin{table}
\[ \begin{array}{|c|c|c|c|}
     \hline
     \mathcal{N} & G & H & \tilde{H}\\
     \hline
     8 & E_7 & S U (8) /\mathbb{Z}_2 & S U (8)\\
     \hline
     6 & S O^{\ast} (12) & U (6) /\mathbb{Z}_2 & U (6)\\
     \hline
     5 & S U (5, 1) & U (5) /\mathbb{Z}_2 & U (5)\\
     \hline
   \end{array} \]
   \caption{Duality groups of $\mathcal{N} \geqslant 4$ supergravities in $4$ dimensions.}
\end{table}

For 16 supersymmetries, in $4<d < 10$ dimensions there is a theory of a
supergravity multiplet coupled to $m$ vector multiplets with duality group
\begin{equation}
  G = S O (n, m) \times \mathbb{R}^+ \label{sonm}
\end{equation}
with $n = 10 - d$. The scalar fields take values in the coset space
\begin{equation}
  \frac{G}{H} = \frac{S O (n, m)}{S O (n) \times S O (m)} \times \mathbb{R}^+
\end{equation}
In 10 dimensions there is a (1,0) theory consisting of (1,0) supergravity
coupled to $r$ (1,0) vector multiplets; here these will taken to be abelian,
so the gauge group is $U (1)^r .$ Dimensionally reducing to $d > 4$ dimensions
on $T^n$ with $n = 10 - d$ gives a theory with duality group (\ref{sonm})
where $m = n + r$.

In $d = 4$, with $n = 6$, the symmetry is enhanced to
\begin{equation}
  G = S O (6, m) \times S L (2, \mathbb{R})
\end{equation}
while in $d = 3$ it is enhanced to
\begin{equation}
  G = S O (8, m + 2)
\end{equation}
Reducing further to $d = 2$ gives a duality group that is affine $S O (8, m +
2) .$

\section{Symmetries and Actions}

\subsection{Scalar field actions}\label{ScalarActions}

Scalar fields $\phi^i (x)$ taking values in a target space $\mathcal{M}$ with
metric $g_{i j}$ are governed by a non-linear sigma model with action
\begin{equation}
  S = \frac{1}{2} \int g_{i j} (\phi) d \phi^i \wedge \ast d \phi^j
\end{equation}
This is invariant under transformations
\begin{equation}
  \delta \phi^i = \alpha^a \xi_a^i \label{fitrans}
\end{equation}
where $\xi_a^i$ are Killing vectors labelled by the index $a$ and $\alpha^a$
are constant parameters. The maximal symmetry group is then the isometry group
of $\mathcal{M}$.

If $\mathcal{M}$ is a coset space $G / H$ it can instead be formulated in
terms of $\mathcal{V} (x) \in G$. The pull-back of the Maurer-Cartan form
$\partial_{\mu} \mathcal{V}$ $\mathcal{V}^{- 1}$ takes values in the Lie
algebra $\mathfrak{g}$ of $G$. The Lie algebra has an orthogonal decomposition
$\mathfrak{g}=\mathfrak{h} \oplus \mathfrak{k}$ where $\mathfrak{h}$ is the
Lie algebra of $H$, so that the Maurer-Cartan form has a decomposition
\begin{equation}
  \partial_{\mu} \mathcal{V}\mathcal{V}^{- 1} = Q_{\mu} + P_{\mu}
\end{equation}
where
\begin{equation}
  Q_{\mu} \in \mathfrak{h}, \quad P_{\mu} \in \mathfrak{k}
\end{equation}
The sigma-model action is then
\begin{equation}
  S = \frac{1}{2} \int d^d x \sqrt{- g} {\rm{Tr} }(P_{\mu} P^{\mu})
  \label{Pact}
\end{equation}
which is invariant under the transformations (\ref{vtrans}), so that it has a
global invariance under the isometry group $G$ and local invariance under $H$.
The 1-form $Q_{\mu}$ is a connection for the local $H$ symmetry and $P_{\mu}$
can be written as the covariant derivative
\begin{equation}
  P_{\mu} = D_{\mu} \mathcal{V}\mathcal{V}^{- 1}, \quad D_{\mu} \mathcal{V}=
  \partial_{\mu} \mathcal{V}- Q_{\mu} \mathcal{V}
\end{equation}
so that
\begin{equation}
  S = \frac{1}{2} \int d^d x \sqrt{- g} {\rm{Tr} }(D_{\mu}
  \mathcal{V}\mathcal{V}^{- 1} D^{\mu}  \mathcal{V}\mathcal{V}^{- 1})
  \label{Pact2}
\end{equation}
If $H$ has an invariant $\Omega$ such that
\begin{equation}
  h^t \Omega h = \Omega
\end{equation}
then
\begin{equation}
  \mathcal{M}=\mathcal{V}^t \Omega \mathcal{V} \label{momega}
\end{equation}
is invariant under local $H$ transformations but transforms under $G$ as
\begin{equation}
  \mathcal{M} \rightarrow g^t \mathcal{M}g
\end{equation}
The matrix $\mathcal{M}$ is sometimes referred to as a generalised metric.
The action (\ref{Pact}) can instead be written in terms of
the generalised metric as
\begin{equation}
  S = \frac{1}{4} \int d^d x \sqrt{- g} { \rm{Tr} }(\partial_{\mu} \mathcal{M}
  \partial^{\mu}  \mathcal{M}^{- 1}) \label{mact}
\end{equation}
If
$H$ is of orthogonal type $\Omega$ and $\mathcal{M}$  are symmetric while if it is of symplectic
type $\Omega$ and $\mathcal{M}$ are antisymmetric.
 If $H$ is instead of unitary type, there is
instead an invariant satisfying
\begin{equation}
  h^{\dag} \Omega h = \Omega
\end{equation}
so that the generealised metric
\begin{equation}
  \mathcal{M}=\mathcal{V}^{\dag} \Omega \mathcal{V}
\end{equation}
is $H$-invariant. 

\subsection{The $p$-form gauge field action}

Consider now a set of $k$ $p$-form gauge fields $A^{\alpha}$ with field
strength $F^{\alpha} = d A^{\alpha}$ labelled by $a = 1, \ldots, k$. The
action is of the form
\begin{equation}
  S = \frac{1}{2} \int \mathcal{N}_{\alpha \beta} (\phi) F^{\alpha} \wedge
  \ast F^{\beta} + \cdots \label{Fact}
\end{equation}
where in general there is a coupling to the scalars $\phi^i .$ Duality
symmetries of the action act linearly on the $p$-form potentials
\[ \delta A^{\alpha} = \alpha^a (L_a)^{\alpha} {}_{\beta} A^{\beta} \]
where $(L_a)^{\alpha}{} _{\beta}$ are some representation of the symmetry $K$
and $K$ acts on the scalars via isometries (\ref{fitrans}). Thus the duality
symmetry $K$of the scalar plus $p$-form system is a subgroup of the isometry
group $G$ and a subgroup of $G L (k, \mathbb{R})$. For this symmetry to be
possible places stringent constraints on the form of $\mathcal{N}_{\alpha
\beta} (\phi)$; see e.g.\ \cite{Hull:1985pq}.

For $d$ even with $p = d / 2 - 1$ the symmetry of the action can be enlarged
to a duality symmetry of the equations of motion. Defining dual field
strengths are specified by
\begin{equation}
  (\ast G)_{\alpha}^{\mu \nu} = - \frac{1}{\sqrt{- g}} \frac{\delta S}{\delta
  F^{\alpha}_{\mu \nu}} \label{Giss}
\end{equation}
For the action (\ref{Fact}),
\begin{equation}
  G_{\alpha} =\mathcal{N}_{\alpha \beta} (\phi) \ast F^{\beta} + \cdots
\end{equation}
The field equations and Bianchi identities are then
\begin{equation}
  d F^{\alpha} = 0, \quad d G_{\alpha} = 0
\end{equation}
and these are invariant under $G L (2 k, \mathbb{R})$ linear transformations
acting on the $2 k$-vector of $p + 1$ forms
\begin{equation}
  \left(\begin{array}{c}
    F^{\alpha}\\
    G_{\alpha}
  \end{array}\right)
\end{equation}
However, for requiring that these be consistent with (\ref{Giss}) further
restricts the symmetry to $S p (2 k, \mathbb{R})$ for $d = 4 r$ or to $O (k,
k)$ for $d = 4 r + 2$ \cite{Tanii:1984zk,Cremmer:1997ct}.  This was shown first for $d = 4$ {in} \cite{Gaillard:1981rj}.

\subsection{11-dimensional supergravity}

11-dimensional gravity was constructed in \cite{Cremmer:1978km}. It is a theory of a graviton
$\mathcal{G}_{{MN}}$, a 3-form gauge field $A_{M N P}$ and a gravitino
$\Psi_M$ which is a 32-component Marjorana fermion and a spacetime vector. The
bosonic part of the action is
\begin{equation}
  S = \int d^{11} x \sqrt{-\mathcal{G}} \mathcal{R}- \frac{1}{2} \int F \wedge
  \ast F + \frac{1}{6} \int F \wedge F \wedge A
  \label{11act}
\end{equation}
where $F = d A$. The dimensional reduction of this theory gives most of the
supergravity theories discussed in this paper.

\subsection{The Trombone}

Under the constant scalings
\begin{equation}
  \mathcal{G}_{M N} \rightarrow \zeta^2 \mathcal{G}_{M N}, \quad A_{M N P}
  \rightarrow \zeta^3 A_{M N P}, \quad \Psi_M \rightarrow \zeta^{\frac{1}{2}}
  \Psi_M \label{trom}
\end{equation}
the 11-dimensional action scales as
\begin{equation}
  S \rightarrow \zeta^9 S
\end{equation}
and as a result the scaling (\ref{trom}) is a symmetry of the equations of
motion, sometimes known as the trombone symmetry. On dimensional reduction
this gives a trombone symmetry in $d$ dimensions, so that all of the
supergravity theories discussed here have such an $\mathbb{R}^+ $trombone
symmetry if the equations of motion. Thus each of the supergravity theories
with duality group $G$ discussed in section 2 in fact has a symmetry $G \times
\mathbb{R}^+$ of the equations of motion. This plays a role in the discussion
of dimensional reduction that follows.

\section{Dimensional Reduction}

Consider a $D = d + n$ dimensional theory dimensionally reduced on a torus
$T^n$ to $d$ dimensions. Let the coordinates be $X^M = (x^{\mu}, y^m)$ where
$x^{\mu}$ are coordinates on the $d$-dimensional spacetime and $y^m$ are
periodic coordinates on the internal $n$-torus. Fields $\varphi (x, y)$ can be
expanded as a sum of Fourier modes $\varphi_n (x)$, giving an infinite tower of
massive Kaluza-Klein fields in $d$ dimensions. If $\varphi$ is massless in
$D$ dimensions, then the $y$-independent zero modes
are massless fields in $d$ dimensions. For a $D$-dimensional supergravity theory, the truncation to
the zero modes gives a $d$-dimensional supergravity and can be seen as arising
from an ansatz in with all fields are independent of $y$. In this truncation,
the structure of the extra dimensions is irrelevant and exactly the same
massless theory can be obtained by replacing $T^n$ with $\mathbb{R}^n$ and
truncating to the $y$-independent subtheory. As we shall see, this is useful
in understanding the symmetries. In this case, the discrete Kaluza-Klein
spectrum would be replaced by a continuum of states without a mass gap, but
this is unimportant for the truncated theory. In what follows, we shall focus
on the dimensional reduction to the massless theory and truncate out all
Kaluza-Klein modes.

\subsection{Pure gravity}\label{PureGrav}

Consider first the case of pure gravity, following e.g.\
\cite{Cremmer:1997ct}. The metric $d s^2 =\mathcal{G}_{M N} d X^M d X^N$ can be taken to be of the
following form, with all components $G_{\mu \nu} (x), h_{m n} (x), A_{\mu}^m
(x)$ depending on $x$ but not $y$:
\begin{equation}
 d s^2 = G_{\mu \nu} d x^{\mu} d x^{\nu} +
  h_{m n} (d y^m + A_{\mu}^m d x^{\mu}) (d y^n + A_{\nu}^n d x^{\nu})	
  \label{metric}
\end{equation}
The reduction of the $D$-dimensional metric then gives a metric $G_{\mu \nu}
(x)$ in $d$-dimensions, $n (n + 1) / 2$ scalar fields $h_{m n}$ arising from
the internal metric on $T^n$ (or $\mathbb{R}^n$) and $A_{\mu}^m (x)$ are $n$
abelian vector fields. \ The scalar fields $h_{m n}$ are the \ moduli of the
torus, specifying the internal geometry, while the $A_{\mu}^m (x)$ are
connections for a (possibly trivial) torus fibration over the $d$-dimensional
spacetime. This metric ansatz has the useful property that
\begin{equation}
  \mathcal{G}= G h
\end{equation}
where
\begin{equation}
  \mathcal{G}= \det (\mathcal{G}_{M N}), \quad G = \det (G_{\mu \nu}), \quad h
  = \det (h_{m n})
\end{equation}
The $D$-dimensional Einstein-Hilbert action
\begin{equation}
  S_D = \frac{1}{\kappa^2} \int d^D X \sqrt{-\mathcal{G}} \mathcal{R}
\end{equation}
becomes, with the ansatz (\ref{metric}),
\begin{equation}
  S_D = \int d^d x d^n y \sqrt{- G} \sqrt{h} \left( R_G + \frac{1}{4}
  \partial_{\mu} h_{m n} \partial^{\nu} h^{m n} \right) + O (A^2)
\end{equation}
where $h^{m n}$ is the inverse of $h_{m n}$ and (for the moment) the terms
depending on the vector fields $A_{\mu}^m$ are suppressed.

The action is invariant under $D$-dimensional diffeomorphisms in the usual
way, but generic transformations will not preserve the form of the metric. For
an internal space which is $\mathbb{R}^n$, the $y$-independence of $h_{m n}$
is preserved by diffeomorphisms generated by vector fields of the form $\xi^m
= \lambda^m {}_n y^n$ for some constant matrix $\lambda^m{} _n$. These generate $G
L (n, \mathbb{R})$ transformations
\begin{equation}
  y^m {\rightarrow y'}^m = (\Lambda^{- 1})^m {}_n y^n, \quad h_{m n} \rightarrow
  h_{p  q} \Lambda^p{} _m \Lambda^q{}  _n
\end{equation}
where $\Lambda = e^{- \lambda}$ is an invertible constant matrix. The
invariance of the action follows from the transformations
\begin{equation}
  \sqrt{h} \rightarrow \sqrt{h} \det (\Lambda), \qquad \int d^n y \rightarrow
  \int d^n y \det (\Lambda^{- 1})
\end{equation}
For $S L (n, \mathbb{R})$ transformations with $\det (\Lambda) = 1$, the two
terms are separately invariant.

The dimensionally reduced action in $d$-dimensions is
\begin{equation}
  S_d = \int d^d x \sqrt{- G} \sqrt{h} \left( R_G + \frac{1}{4} \partial_{\mu}
  h_{m n} \partial^{\nu} h^{m n} \right) + O (A^2) \label{acth}
\end{equation}
It is invariant under $S L (n, \mathbb{R})$ transformations with $\det
(\Lambda) = 1$, but not under the full $G L (n, \mathbb{R})$ as the
integration over $y$ is missing. Indeed,
\begin{equation}
  S_d \rightarrow \det (\Lambda) S_d \label{scaled}
\end{equation}
so that under the scaling
\begin{equation}
  h_{m n} \rightarrow \alpha^2 h_{m n} \label{hscal}
\end{equation}
the action scales as
\begin{equation}
  S_d \rightarrow \alpha^n S_d \label{scaled}
\end{equation}
This scale transformation is not a symmetry of the action but is a symmetry of
the equations of motion. However, the trombone symmetry in $D$-dimensions
$\mathcal{G}_{M N} \rightarrow \zeta^2 \mathcal{G}_{M N}$ gives the symmetry
\begin{equation}
  G_{\mu \nu} \rightarrow \zeta^2 G_{\mu \nu}, \quad h_{m n} \rightarrow
  \zeta^2 h_{m n}, \quad A_{\mu}^m \rightarrow A_{\mu}^m \label{tromD}
\end{equation}
under which
\[ S_d \rightarrow \zeta^{D - 2} S_d \]
and so is a symmetry of the equations of motion. Thus the symmetry of the
equations of motion of $S_d$ is
\begin{equation}
  G L (n, \mathbb{R}) \times \mathbb{R}^+ \label{glnr}
\end{equation}
(Here and throughout, $\mathbb{R}^+ $ is the multiplicative group of positive
real numbers, isomorphic to the additive group of real numbers $\mathbb{R}.$)
However the $G L (n, \mathbb{R})$ subgroup in which
\begin{equation}
  \zeta^{2 - D} = \det (\Lambda)
\end{equation}
is a symmetry of the action, consisting of $S L (n, \mathbb{R})$
transformations together with the scale transformation
\begin{equation}
  G_{\mu \nu} \rightarrow \rho^2 G_{\mu \nu}, \quad h_{m n} \rightarrow
  \rho^{- 2 \alpha} h_{m n}, \quad A_{\mu}^m \rightarrow A_{\mu}^m
\end{equation}
with
\begin{equation}
  \alpha = \frac{d - 2}{n}
\end{equation}
Thus the $G L (n, \mathbb{R})$ symmetry arises from diffeomorphisms of
$\mathbb{R}^n$ together with a trombone transformation. There is a further
$\mathbb{R}^+ $ symmetry of the equations of motion which could be taken to be
the $D$-dimensional trombone symmetry $\mathcal{G}_{M N} \rightarrow \zeta^2
\mathcal{G}_{M N}$ giving (\ref{tromD}) or (combining this with (\ref{hscal}))
to be the $d$-dimensional trombone symmetry
\begin{equation}
  G_{\mu \nu} \rightarrow \sigma^2 G_{\mu \nu}, \quad h_{m n} \rightarrow h_{m
  n}, \quad A_{\mu}^m \rightarrow A_{\mu}^m
\end{equation}
For the case in which the internal space is $T^n$ rather than $\mathbb{R}^n$
with periodic coordinates $y^m \sim y^m + 1$, the vector fields of the form
$\xi^m = \lambda^m {}_n y^n$ are not well-defined on $T^n$. In this case the
transformation $y^m {\rightarrow y'}^m = (\Lambda^{- 1})^m {}_n y^n$ is only
consistent with the periodicity condition if $(\Lambda^{- 1})^m {}_n$ is an
integer-valued matrix, so that the only allowed diffeomorphisms are those in
$G L (n, \mathbb{Z})$; these are the large diffeomorphisms of the torus,
forming the mapping class group. Nonetheless, there is a $G L (n, \mathbb{R})$
action on the internal metric $h_{m n} \rightarrow h_{p  q} \Lambda^p{}
_m \Lambda^q {} _n$ which is well-defined for any invertible matrix $\Lambda^p{}
_m$. Thus $S_d$ has a $G L (n, \mathbb{R})$ symmetry while $S_D$ only has a $G
L (n, \mathbb{Z})$ diffeomorphism invariance. Thus a general $G L (n,
\mathbb{R})$ transformation in $d$ dimensions does not lift to a
diffeomorphism, corresponding to the vector field $\xi^m = \lambda^m {}_n y^n$
being only locally defined. 
Such transformations can be viewed as generalised symmetries; see e.g.\ \cite{Gomez-Fayren:2024cpl}.
The key point is that $\delta g_{mn}= 2\nabla_{( m}\xi_{n)}$ can be a globally-defined tensor for certain $\xi^m$ that are only locally defined.

The metric can be written in terms of a vielbein $\mathcal{V}^a_m (x) = e^a_m$
with
\begin{equation}
  h_{m n} = \delta_{a b} e^a_m e^b_n \label{vielh}
\end{equation}
and this is invariant under local $O (n)$ transformations $\mathcal{V}^a_m
\rightarrow M^a{} _b (x) \mathcal{V}^b_m$. The $n \times n$ matrix
$\mathcal{V}^a_m$ is invertible and so can be viewed as an element of $G L (n,
\mathbb{R}) .$ The metric is defined by gauge orbits of $\mathcal{V}^a_m$
under the action of local $O (n)$ transformations and so can be identified
with an element of the coset
\begin{equation}
  \frac{G}{H} = \frac{G L (n, \mathbb{R})}{O (n)} \label{ghnr}
\end{equation}
The $n (n + 1) / 2$ scalar fields take values in this target space.

The transformations in $G L (n, \mathbb{R})$ with $\det (L^m {} _n)<0 $ change
the orientation of the internal torus. For example the transformation $y^1
\rightarrow - y^1$ is a reflection in the $y^1$ direction that leads to the
symmetry of the reduced action under $A_{\mu}^1 \rightarrow - A_{\mu}^1$ while
other components of $A_{\mu}^m$ are invariant. Such transformations lead to
complications for theories involving fermions -- for example they would need
to be represented by pinor fields rather than spinor fields. It will be
convenient to fix an orientation for the internal space and restrict the
transformations to be in the subgroup $G L (n, \mathbb{R})^+$ of $G L (n,
\mathbb{R})$ with $\det (L^m {} _n)>0$ that preserve that orientation,
\[ G L (n, \mathbb{R})^+ = S L (n, \mathbb{R}) \times \mathbb{R}^+ \]
The scalar coset can then be viewed as
\begin{equation}
  \frac{G}{H} = \frac{G L (n, \mathbb{R})^+}{S O (n)} = \frac{S L (n,
  \mathbb{R}) }{S O (n)} \times \mathbb{R}^+
\end{equation}
Writing
\begin{equation}
  h_{m n} = e^{- 4 \Phi / n} \bar{h}_{m n}
\end{equation}
with
\begin{equation}
  \det (\bar{h}_{m n}) = 1, \quad e^{- 4 \Phi} = \det (h_{m n})
\end{equation}
the action (\ref{acth}) can be written as
\begin{equation}
  S_d = \int d^d x \sqrt{- G} e^{- 2 \Phi} \bigg( R_G + \frac{1}{4}
  \partial_{\mu} \bar{h}_{m n} \partial^{\nu} \bar{h}^{m n}   
   + \frac{\alpha}{2}
  (\partial \Phi)^2  \bigg) + O (A^2)	
\end{equation}
so that the $\mathbb{R}^+$ symmetry acts as a shift of the ``dilaton'' field
$\Phi$:
\begin{equation}
  \Phi \rightarrow \Phi + X, \quad G_{\mu \nu} \rightarrow e^{- X} G_{\mu
  \nu}, \quad \bar{h}_{m n} \rightarrow \bar{h}_{m n} \label{dilsh}
\end{equation}
A Weyl rescaling gives a metric
\begin{equation}
  g_{\mu \nu} = e^{\Phi} G_{\mu \nu}
\end{equation}
which is invariant under (\ref{dilsh}) and takes the full action to the Einstein
frame action
\begin{equation}
  S_d = \int d^d x \sqrt{- g} \bigg( R_g - \frac{1}{4} \partial_{\mu}
  \bar{h}_{m n} \partial^{\nu} \bar{h}^{m n} - \frac{\alpha}{2} (\partial
  \Phi)^2   - \frac{1}{4} e^{2 \alpha \Phi} \bar{h}_{m n} F^m_{\mu \nu} F^{\mu
  \nu n} \bigg)	
\end{equation}

with
\[ \alpha = \frac{D - 2}{n (d - 2)} \]
The symmetry of the action remains $G L (n, \mathbb{R})$ while the
symmetry of the equations of motion is $G L (n, \mathbb{R}) \times
\mathbb{R}^+ .$

\subsection{Gravity with a $p$-form gauge field}

Consider now gravity coupled to a $p$-form gauge field $A$ with field strength
$F = d A$, with the following action added to the Einstein-Hilbert action:
\begin{equation}
  S = \frac{1}{2} \int F \wedge \ast F
\end{equation}
The gauge field $A$ is taken to depend only on $x^{\mu}$ and to be independent
of the internal coordinates $y^m$; see e.g.\ \cite{Samtleben:2023ivs}for details. Compactifying on
$T^n$ (or $\mathbb{R}^n$ as discussed above), gives rise to a set of $q$-form
gauge fields in $d$ dimensions $A_{\mu_1 \ldots \mu_q m_{q + 1} \ldots m_p}$
with $p - n \leqslant q \leqslant p$, with the number of $q$-forms given by
the binomial coefficient $\left(\begin{array}{c}
  n\\
  q
\end{array}\right)$. In particular, if $p \leqslant n$ there are
$\left(\begin{array}{c}
  n\\
  p
\end{array}\right)$ scalar fields arising from the internal components $A_{m_1
\ldots m_p}$. The $G L (n, \mathbb{R})$ diffeomorphism symmetry and the
$\mathbb{R}^+$ trombone symmetry act on these fields, so that the $G L (n,
\mathbb{R}) \times \mathbb{R}^+$ symmetry of pure gravity extends to a
symmetry of gravity coupled to a $p$-form gauge field dimensionally reduced
from $D$ to $d = D - n$ dimensions.

The theory in $D$ dimensions has the $p$-form symmetry
\begin{equation}
  \delta A = \zeta, \quad d \zeta = 0
\end{equation}
corresponding to shifts by a closed $p$-form $\zeta$. Closed forms $\zeta = d
\lambda$ that are exact give the gauge symmetry
\begin{equation}
  \delta A = d \lambda
\end{equation}
with $p - 1$ form parameter $\lambda$, while shifts by closed forms modulo
exact forms define global symmetries, sometimes referred to as generalised
symmetries. In particular, for $p \leqslant n$ the shifts of the internal
components $A_{m_1 \ldots m_p}$ by constants $\zeta_{m_1 \ldots m_p}$
\begin{equation}
  \delta A_{m_1 \ldots m_p} = \zeta_{m_1 \ldots m_p}
\end{equation}
preserves the the ansatz that $A$ is independent of $y$ and so gives a
symmetry in $d$ dimensions that shifts the scalars $A_{m_1 \ldots m_p}$ by
constants, so that these scalars have an axionic symmetry. For an internal
space that is $\mathbb{R}^n$ these are gauge transformations with
$\lambda_{m_1 \ldots m_{p - 1}} = \zeta_{m_1 \ldots m_p} y^{m_p}$ while for an
internal space that is $T^n$ these are $p$-form generalised symmetries  \cite{Gomez-Fayren:2024cpl}.

Then the theory of gravity coupled to a $p$-form gauge field has a
symmetry
\begin{equation}
  G L (n, \mathbb{R}) \times \mathbb{R}^+ \ltimes \mathbb{R}^N, {} \quad
  N = \left(\begin{array}{c}
    n\\
    p
  \end{array}\right)
\end{equation}
consisting of diffeomorphisms, the trombone symmetry and the axionic shifts. A
subgroup
\begin{equation}
  G L (n, \mathbb{R}) \ltimes \mathbb{R}^N
\end{equation}
is a symmetry of the action. In particular, 11-dimensional gravity has a
bosonic sector consisting of gravity coupled ot a 3-form gauge field, so that
compactification on $T^n$ gives an action with symmetry
\begin{equation}
  G L (n, \mathbb{R}) \ltimes \mathbb{R}^{n (n - 1) (n - 2) / 6}
\end{equation}
for $n \geqslant 3.$

\subsection{Gravity Coupled to a 2-form gauge field and a dilaton}
\label{gbf}

Consider the $D$-dimensional action
\begin{equation}
  S = \int \sqrt{-\mathcal{G}} e^{- 2 \Phi} \left( \mathcal{R}- 2 (\partial
  \Phi)^2 - \frac{1}{6} H^2 \right) \label{Bact}
\end{equation}
where $H = d B$ is the 3-form field strength for a 2-form gauge field $B$. For
$D = 10$, this is the bosonic sector of (1,0) supergravity and is a subsector
of the action of the $(2, 0)$ and $(1, 1)$ supergravities, which are the
low-energy effective actions for {IIB} and {IIA} superstrings,
respectively. The dimensional reduction on  $T^n$ was analysed in \cite{Maharana:1992my} and gives a theory of a metric
$g_{\mu \nu}$, $2 n$ vector fields $A_{\mu}^m, B_{\mu m}$ and $n^2 + 1$
scalars $h_{m n}, B_{m n}, \Phi$ and a 2-form $B_{\mu \nu}$.

The action for the scalars $h_{m n}, B_{m n}$ in $d$ dimensions can be
written in matrix notation as
\begin{equation}
  S = \frac{1}{4} \int \sqrt{- g} {\rm{Tr} } (\partial_{\mu} h^{- 1}
  \partial^{\mu} h + h^{- 1} \partial_{\mu} B h^{- 1} \partial^{\mu} B)
\end{equation}
The first term is the scalar term in (\ref{acth}). Introducing the $2 n \times
2 n$ matrices with $n \times n$ blocks
\begin{equation}
  \mathcal{M}= \left(\begin{array}{cc}
    h^{- 1} & - h^{- 1} B\\
    B h^{- 1} & h - B h^{- 1} B
  \end{array}\right) \label{miss}
\end{equation}
and the $O (n, n)$-invariant metric
\begin{equation}
  \eta = \left(\begin{array}{cc}
    0 & 1\\
    1 & 0
  \end{array}\right)
\end{equation}
the action can be written in the form (\ref{mact}):
\begin{equation}
  S = \frac{1}{8} \int \sqrt{- g} \, {\rm{Tr}} (\partial_{\mu} \mathcal{M}^{- 1}
  \partial^{\mu} \mathcal{M})
\end{equation}
This is invariant under the rigid $O (n, n)$ transformations
\begin{equation}
  \mathcal{M} \rightarrow \Lambda^t \mathcal{M} \Lambda \label{mtrans}
\end{equation}
where the matrix $\Lambda \in O (n, n)$ satisfies
\begin{equation}
  \Lambda^t \eta \Lambda = \eta \label{onncon}
\end{equation}
The action can also be written in terms of a vielbein
\[ \mathcal{V}= \left(\begin{array}{cc}
     e^{- 1} & - e^{- 1} B\\
     0 & e
   \end{array}\right) \]
where from (\ref{vielh}) $h = e^t e$. Then
\begin{equation}
  \mathcal{M}=\mathcal{V}^t \mathcal{V}
\end{equation}
which is of the form (\ref{momega}) with $\Omega =\mathbb{I}$ the $O (n)
\times O (n)$ invariant metric. The vielbein is an element of $O (n, n)$ as
\begin{equation}
  \mathcal{V}^t \eta \mathcal{V}= \eta
\end{equation}
The matrix $\mathcal{M}=\mathcal{V}^t \mathcal{V}$ is invariant under local $O
(n) \times O (n) $transformations $\mathcal{V} \rightarrow h\mathcal{V}$ while
under rigid $O (n, n) $ transformations $\mathcal{V} \rightarrow \mathcal{V}
\Lambda$ it transforms as (\ref{mtrans}). The action can then be written in
the form (\ref{Pact2}). Then the discussion of section \ref{ScalarActions}
applies and $\mathcal{V}$ or $\mathcal{M}$ parameterise the coset space
\begin{equation}
  \frac{O (n, n)}{O (n) \times O (n)}
\end{equation}
As discussed in section \ref{PureGrav}, it is convenient to restrict to
orientation-preserving transformations and take the scalar coest space to be
\begin{equation}
  \frac{S O (n, n)}{S O (n) \times S O (n)}
\end{equation}
Then the scalars parameterise $h_{m n}, B_{m n}, \Phi$ then the coset space
\begin{equation}
  \frac{S O (n, n)}{S O (n) \times S O (n)} \times \mathbb{R}^+
  \label{sonn}
\end{equation}
The full action has an $S O (n, n)$ duality symmetry under which the $2 n$
vector fields transform in the $\mathbf{2 n}$ vector representation \cite{Maharana:1992my}.
This contains the expected $G L (n, \mathbb{R})$ symmetry from diffeomorphisms
and the $\mathbb{R}^{n (n - 1) / 2}$ symmetry from shifts of the axions $B_{m
n}$ and the $\mathbb{R}^+$ from shifts of the dilaton. However, there are also
transformations that mix the metric components  $h_{m n}$ with the $B$-field
components $B_{m n}$ and such transformations do not have any interpretation
as a symmetry in $D$ dimensions. As usual, the field equations have a further
$\mathbb{R}^+ $trombone symmetry.

\section{Hidden Symmetry}

\subsection{Ehlers Symmetry: $S L (2, \mathbb{R})$ duality of $\mathcal{N}= 1$
$D = 4$ supergravity  reduced to 3 dimensions} \label{Ehlers}

From the previous section, dimensionally reducing pure gravity in $D = 4$
dimensions to $d = 3$ dimensions gives a theory with a metric $g_{\mu \nu}$, a
vector field $A_{\mu}$ and a scalar $\Phi$. The lagrangian in 3-dimensions is
\begin{equation}
  L = \sqrt{- g} \left( R - \frac{1}{2} (\partial \Phi)^2 - \frac{1}{4} e^{- 2
  \Phi} F^2 \right) \label{lagr}
\end{equation}
This has an $\mathbb{R}^+$ symmetry
\begin{equation}
  \Phi \rightarrow \Phi + \sigma, \quad A_{\mu} \rightarrow e^{\sigma} A_{\mu}
\end{equation}
for constant parameter $\sigma .$

In 3 dimensions, a vector field is dual to a scalar field so that $A_{\mu}$
can be dualised to give a scalar $\chi$, and the dual action is
\begin{equation}
  L = \sqrt{- g} \left( R - \frac{1}{2} (\partial \Phi)^2 - \frac{1}{2} e^{2
  \Phi} (\partial \chi)^2 \right) \label{dualact}
\end{equation}
This can be seen as follows. In (\ref{lagr}), the 2-form $F = d A$ satisfies
$d F = 0$. Instead, the action can be written in terms of an unconstrained $F$
with $d F = 0$ imposed by adding a Lagrange multiplier term
\begin{equation}
  L = \sqrt{- g} \left( R - \frac{1}{2} (\partial \Phi)^2 - \frac{1}{4} e^{- 2
  \Phi} F^2 + \frac{1}{6} \chi \varepsilon^{\mu \nu \rho} \partial_{\mu}
  F_{\nu \rho} \right)
\end{equation}
Integrating over $\chi$ imposes the constraint $d F = 0$ so that the original
theory (\ref{lagr}) is recovered. On the other hand, integrating out $F$ gives
the dual action (\ref{dualact}). As both are obtained from the same parent
theory, they are physically equivalent.

The action (\ref{dualact}) gives gravity coupled to 2 scalar fields. The
action for the scalars is that of a non-linear sigma-model with a
2-dimensional target space with metric
\begin{equation}
  d s^2 = d \Phi^2 + e^{2 \Phi} d \chi^2
\end{equation}
This is the metric on the coset space $S L (2, \mathbb{R}) / U (1)$, so that
the theory is gravity coupled to a sigma model with target space $S L (2,
\mathbb{R}) / U (1)$. The isometry group of this coset space is $S L (2,
\mathbb{R})$ and this gives a symmetry of (\ref{dualact}) under which
\begin{equation}
  \tau = \chi + i e^{i \Phi}
\end{equation}
transforms by fractional linear transformations
\begin{equation}
  \tau \rightarrow \frac{a \tau + b}{c \tau + d} \label{fracl}
\end{equation}
where the 4 real numbers $a, b, c, d$ satisfy
\begin{equation}
  a d - b c = 1
\end{equation}
so that
\begin{equation}
  \left(\begin{array}{cc}
    a & b\\
    c & d
  \end{array}\right) \in S L (2, \mathbb{R})
\end{equation}
This  is the {\it Ehlers symmetry} \cite{Ehlers:1959aug,Geroch:1972yt}.
This symmetry is generated by constant axionic shifts $\chi \rightarrow \chi +
c$, dilaton shifts $\Phi \rightarrow \Phi + \sigma$, $\chi \rightarrow e^{-
\sigma} \chi$ and a non-linear inversion $\tau \rightarrow - 1 / \tau$. The
two shift symmetries are to be expected but the non-linear inversion symmetry
is rather surprising and such unexpected symmetries are sometimes referred to
as ``hidden symmetries''. The $S L (2, \mathbb{R})$ symmetry is the Ehlers
symmetry. $\mathcal{N}= 1$ supergravity in $D = 4$ is Einstein's theory
coupled to a gravitino field, and its reduction to $d = 3$ also has $S L (2,
\mathbb{R})$ duality symmetry and scalars in the coset space $S L (2,
\mathbb{R}) / U (1)$.

\subsection{Gravity in $D = 3 + n$ dimensions reduced to 3
dimensions}\label{SLN}

Consider now gravity in $D = 3 + n$ dimensions reduced to $d = 3$ dimensions.
This gives a theory with a metric $g_{\mu \nu}$, $n$ vector fields $A^m_{\mu}$
and $n (n + 1) / 2$ scalars $G_{m n}$ and has a $G L (n, \mathbb{R})$
symmetry. Dualising the vectors gives a further $n$ scalars $\chi^m$. The
dualised theory is gravity coupled to a non-linear sigma model whose target
space has dimension $n (n + 1) / 2 + n = (n + 1) (n + 2) / 2 - 1$.
Compactifying first to 4 dimensions gives a $G L (n - 1, \mathbb{R})$
symmetry, then further reducing to $d = 3$ gives a further $S L (2,
\mathbb{R})$ Ehlers symmetry, so that there should be at least 
 $G L (n - 1,
\mathbb{R}) \times S L (2, \mathbb{R})$ symmetry. This does not commute with
the $G L (n, \mathbb{R})$ symmetry that follows from reducing from $D = 3 + n$
dimensions to $3$ dimensions in one go, so the symmetry should be a larger
group that contains the group generated by the $G L (n - 1, \mathbb{R}) \times
S L (2, \mathbb{R})$ and \ $G L (n, \mathbb{R})$ symmetries. In this case the
duality group is $S L (n + 1, \mathbb{R})$ and the target space is the coset
space
\begin{equation}
  \frac{S L (n + 1, \mathbb{R})}{S O (n + 1)}
\end{equation}
\subsection{Harrison Symmetry: the $S U (2, 1)$ duality of $\mathcal{N}= 2$ $D
= 4$ supergravity  reduced to 3 dimensions }\label{harri}

Consider next Einstein-Maxwell theory in $D = 4$ reduced to $d = 3$.
Dimensionally reducing, 
the metric $\mathcal{G}_{M
N}$ gives a metric $g_{\mu \nu}$, a vector field $A_{\mu}$ and a scalar $\Phi$
while the vector field $\mathcal{C}_M$ gives a vector field $C_{\mu}$ and a
scalar $C_3$. The theory has a symmetry (\ref{glnr}) which in this case is $G
L (1, \mathbb{R}) \ltimes \mathbb{R}^+$. The two vector fields $A_{\mu},
C_{\mu}$ dualise to give two scalars, so that there is a total of four
scalars. The $G L (1, \mathbb{R})$ is enhanced to at least the Ehlers group $S
L (2, \mathbb{R})$. The symmetry of the theory is in fact $S U (2, 1)$ and the
4 scalars parameterise the coset space
\begin{equation}
  \frac{S U (2, 1)}{S (U (2) \times U (1))}
\end{equation}
The $S U (2, 1)$ invariance is often referred to as  {\it Harrison symmetry}.
$\mathcal{N}= 2$ supergravity in $D = 4$ is Einstein-Maxwell theory coupled to
two gravitino fields, and its reduction to $d = 3$ also has $S U (2, 1)$
duality symmetry and scalars in the coset space $\frac{S U (2, 1)}{S (U (2)
\times U (1))}$.

$D = 4$ gravity coupled to $r$ vector fields has an $S O (r)$ symmetry and
reduction to $d = 3$ gives a theory with $r + 1$ vector fields and $r + 1$
scalars, so that dualising the vectors gives a theory with $2 (r + 1)
$scalars. For each vector, there should be an $S U (2, 1)$ symmetry and there
is an $S O (r)$ symmetry rotating the vectors into each other and together
these generate th symmetry group $S U (r + 1, 1)$ and the scalars take values
in the coset
\begin{equation}
  \frac{S U (r + 1, 1)}{S (U (r + 1) \times U (1))}
\end{equation}
which indeed has dimension $2 (r + 1)$.

\subsection{$G_2$ duality of $\mathcal{N}= 2$ $D = 5$ supergravity reduced to
3 dimensions}

As another example, $\mathcal{N}= 2$ supergravity in $D = 5$ dimensions (with
8 supersymmetries) has a bosonic sector consisting of a metric $\mathcal{G}_{M
N}$ and a vector field $\mathcal{A}_M$ with field strength $\mathcal{F}=
d\mathcal{A}$. The action is
\begin{equation}
  S = \int \mathcal{R} \ast 1 - \frac{1}{2} \mathcal{F} \wedge \ast
  \mathcal{F}+ \frac{1}{3 \sqrt{3}} \mathcal{F} \wedge \mathcal{F} \wedge
  \mathcal{A}
\end{equation}
has a Chern-Simons term. Dimensionally reducing to $d = 3$ gives a theory with
a metric $g_{\mu \nu}$ and three vector fields $A^m_{\mu}, \mathcal{A}_{\mu}$
together with 5 scalars $G_{m n}, \mathcal{A}_m$. Dualising the vectors gives
8 scalars and these parameterise the coset space
\begin{equation}
  \frac{G_2}{S U (2) \times 
  {SU} (2)}
\end{equation}
and the duality symmetry is $G_2$ \cite{Mizoguchi:1998wv,Cremmer:1999du}. (This is the non-compact form of $G_2$
whose maximal compact subgroup is $S U (2) \times {SU} (2)$.) This is the
duality group of the $d = 3$ supergravity obtained from dimensionally reducing
$\mathcal{N}= 2$ supergravity in $D = 5$ dimensions.

\subsection{S-duality: the $S L (2, \mathbb{R})$ duality of gravity in $D$
dimensions coupled to $D - 2$ form gauge field}\label{SL2}

Consider the action in $D$ dimensions for a metric $g_{\mu \nu}$, a $D - 2$
form gauge field $A_{\mu_1 \ldots \mu_{D - 2}}$ with field strength $F = d A$
and a scalar $\Phi$ with lagrangian \
\begin{equation}
  L = \sqrt{- g} \left( R - \frac{1}{2} (\partial \Phi)^2 - \frac{1}{2 (D - 1)
  !} e^{- 2 \Phi} F^2 \right)
\end{equation}
For $D = 3$, this is the lagrangian (\ref{lagr}) and has an $\mathbb{R}^+$
symmetry
\begin{equation}
  \Phi \rightarrow \Phi + \sigma, \quad A  \rightarrow e^{\sigma} A 
\end{equation}
for constant parameter $\sigma .$

In $D$ dimensions, a $D - 2$ form gauge field is dual to a scalar field so
that $A $ can be dualised to give a scalar $\chi$, and the dual action is
again

\begin{equation}
  L = \sqrt{- g} \left( R - \frac{1}{2} (\partial \Phi)^2 - \frac{1}{2} e^{2
  \Phi} (\partial \chi)^2 \right) \label{qwe}
\end{equation}
by an argument similar to the one given above for $D = 3$. The lagrangian
(\ref{qwe}) gives gravity in $D$ dimensions coupled to 2 scalar fields taking
values in the coset space $S L (2, \mathbb{R}) / U (1)$, and the theory again
has the duality symmetry ${SL} (2, \mathbb{R})$ acting on $\tau = \chi +
i e^{i \Phi}$ by the fractional linear transformation (\ref{fracl}).

\subsection{$(1, 0)$ supergravity in $D = 10$ \ reduced to $d$ dimensions}

$(1, 0)$ supergravity in $D = 10$ has a metric $\mathcal{G}_{M N}$, a 2-form
gauge field $B_{M N}$ and a dilaton with action (\ref{Bact}) for $D = 10$.
Reducing on $T^n$ gives a theory of a metric $g_{\mu \nu}$, $2 n$ vector
fields $A_{\mu}^m, B_{\mu m}$ and $n^2 + 1$ scalars $G_{m n}, B_{m n}, \Phi$
and a 2-form $B_{\mu \nu}$. These fit into a supergravity multiplet plus $n$
abelian vector multiplets The scalars parameterise the coset space (\ref{sonn})
and there is a $S O (n, n) \times \mathbb{R}^+$ duality symmetry of the
action. For the supergravity in $10 - n$ dimensions coupled to $m$ vector
multiplets this is replaced by
\begin{equation}
  \frac{S O (n, m)}{S O (n) \times S O (m)} \times \mathbb{R}^+ \label{sonmaa}
\end{equation}
with $S O (n, m) \times \mathbb{R}^+$ duality symmetry. For $(1, 0)$
supergravity in $D = 10$ coupled to $r$ $(1, 0) $ abelian vector multiplets
(with gauge group $U (1)^r$), dimensionally reducing to $d = 10 - n$ gives a
supergravity multiplet plus $m = n + r$ abelian vector multiplets. This is the
effective supergravity theory for the heterotic string compactified on $T^n$
at a generic point in the moduli space at which the gauge symmetry is broken
to an abelian subgroup.

In $d = 4$ the 2-form $B$ can be dualised to a scalar $\chi$ so that, as in
section \ref{SL2}, the scalar fields $\Phi, \chi$ are governed by a $S L (2,
\mathbb{R}) / U (1)$ sigma model. As a result, the scalar sigma model for the
full theory has target space
\begin{equation}
  \frac{S O (6, 6)}{S O (6) \times S O (6)} \times \frac{S L (2,
  \mathbb{R})}{U (1)} \label{cost}
\end{equation}
The $12$ vector fields $\mathcal{A}_{\mu}^I = (A_{\mu}^m, B_{\mu m})$ (with $I
= 1, \ldots, 12$ labelling the vector representation of $O (6, 6)$ have field
strengths $\mathcal{F}^I_{\mu \nu}$. The dual field strengths are specified by
\begin{equation}
  (\ast \mathcal{G})_I^{\mu \nu} = - \frac{1}{\sqrt{- g}} \frac{\delta
  S}{\delta \mathcal{F}^I_{\mu \nu}} \label{gdust}
\end{equation}
The $S L (2, \mathbb{R})$ symmetry of the scalar sector extends to a symmetry
of the full equations of motion with the $S L (2, \mathbb{R})$ acting through
S-duality, with the pair $(\mathcal{F}^I_{\mu \nu}, \mathcal{G}_I^{\mu \nu})$
transforming as an $S L (2, \mathbb{R})$ doublet for each $I$. Then the
duality symmetry is
\begin{equation}
  G = S O (6, 6) \times S L (2, \mathbb{R})
\end{equation}
acting as isometries of the coset space (\ref{cost}) and with the field
strengths $(\mathcal{F}^I_{\mu \nu}, \mathcal{G}_I^{\mu \nu})$ transforming in
the $\mathbf{(12, 2)}$ representation.

For $(1, 0)$ supergravity in $D = 10$ coupled to $r$ $(1, 0) $ abelian vector
multiplets (with gauge group $U (1)^r$), dimensionally reducing to $d = 4$
gives a supergravity multiplet plus $m = 6 + r$ abelian vector multiplets for
$\mathcal{N}= 4$ supergravity in $d = 4.$ In this case the coset space is
\begin{equation}
  \frac{S O (6, m)}{S O (6) \times S O (m)} \times \frac{S L (2,
  \mathbb{R})}{U (1)}
\end{equation}
and the duality group is
\begin{equation}
  G = S O (6, m) \times S L (2, \mathbb{R})
\end{equation}
with compact subgroup $H = S O (6) \times S O (m) \times U (1)$. The fermions
transform under a (partial) double cover of this:
\begin{equation}
  \tilde{H} = {Spin} (6) \times S O (m) \times U (1)
\end{equation}
Reduction to $d = 3$ gives a duality symmetry $S O (7, 7 + r) \times
\mathbb{R}^+$. This gives a theory with $7 (7 + r) + 1$ scalars and $14 + r$
vector fields. However, the vector fields can be dualised to give a further 
$14 + r$ scalars, so that the total number of scalars is $8 (8 + r)$.

From section \ref{SLN}, for reduction to $d = 3$, the $G L (7, \mathbb{R})$
subgroup of $S O (7, 7)$ is enhanced to $S L (8, \mathbb{R})$, while from
section \ref{harri} there is an $S U (r + 1, 1)$ symmetry, and as these do not
commute they must generate a larger symmetry group. The full symmetry group is
in fact $S O (8, 8 + r)$ with the scalar fields taking values in
\begin{equation}
  \frac{S O (8, 8 + r)}{S O (8) \times S O (8 + r)}
   \label{cost8}
\end{equation}
which indeed has dimension $8 (8 + r)$.

\section{Duality symmetry of maximal supergravity}

The aim now is to understand the $E_n $ duality symmetry in $d = 11 - n$ dimensions
for the of maximal supergravity theory in which all fields are dualised (where possible) to the lowest
possible rank; there is a further $\mathbb{R}^+ $  trombone symmetry of the
equations of motion in each case. 
The Lie algebra of $G$ can be determined by the following \lq subgroup argument'.
For dimensions $d \leqslant 9$ the
(ungauged) maximal supergravity theory is unique and can be obtained by
reducing 11-dimensional gravity on $T^{n}$ 
so that there is   a $G L (n, \mathbb{R})$ symmetry associated with
diffeomorphisms, and in 9 dimensions $E_2=G L (2, \mathbb{R})$ is the full duality group.
Proceeding inductively, if it is assumed that there is a
duality symmetry $E_{n - 1}$ from reduction on $T^{n-1}$, then there will be at least a
symmetry $E_{n - 1} \times \mathbb{R}^+$ from reduction on $T^{n}$. 
The two rank $n$ groups $G L (n, \mathbb{R})$ and $E_{n - 1} \times \mathbb{R}^+$ do not commute
and together generate a  group $G$ of rank at least $n$, and this group must itself be a duality symmetry.
However, these two groups are both  subgroups of $E_n$ so the group $G$ generated this way must be 
a subgroup of $E_n$, $G\subseteq E_n$. Moreover, they are both maximal subgroups, so the duality group should be  
$G=E_n$. This argument does not fix the global form of the group e.g.\ for $n=5$, it doesn't determine whether the duality group is $SO(5,5)$ or $Spin(5,5)$ or $O(5,5)$ etc. In this case, the 16 vector fields fit into a spinor representation, so that the group is (at least) $G=Spin(5,5)$.

 Both IIA and IIB supergravities have a common sector consisting of a metric, $B$-field and dilaton and 
 it was seen in section \ref{gbf} that compactifying  this sector on an $n-1$ torus  to $d$ dimensions gives 
  a theory with $SO(n-1,n-1)$ symmetry. It is interesting to ask whether this extends to the full theory.
  In fact the duality group $E_n$ has a subgroup $Spin(n-1,n-1)$ under which the remaining bosonic fields (arising from the RR sector of string theory) transform as a spinor under this group \cite{Hull:1994ys}, which is often referred to as the T-duality group. In fact, $E_n$ has a maximal subgroup 
 $Spin(n-1,n-1)\times \mathbb{R}^+$ for $n<7$,  $E_7$ has a maximal subgroup 
 $Spin(6,6)\times SL(2,\mathbb{R})$ (the product of T and S dualities) and $E_8$ has a maximal subgroup 
 $Spin(8,8)$.
 
There is a different $E_{n - 1} \times \mathbb{R}^+$ symmetry group arising  for each choice of $T^{n-1}$ in $T^{n}$; these do not commute with each other and are mapped into each other by the action of $G L (n, \mathbb{R})$.
The group $E_n$ also contains the group 
\begin{equation}
  S L (2, \mathbb{R}) \times S L (n - 1, \mathbb{R}) 
\end{equation}
expected from reduction of IIB supergravity on $T^{n-1}$
In $d \leqslant 6$ there is  Ehlers-type symmetry
enhancement arising from dualisations:
dualisations of $d - 2$ form gauge fields to
scalars yields an $S L (2, \mathbb{R})$ symmetry for each $d - 2$ form, as
seen in section \ref{SL2}.
This will now be explored in various
dimensions.


\subsection{9 and 10 dimensions and T-duality}

In $D = 10,$ there are 2 supergravities with 32 supersymmetries, the IIA
theory with (1,1) supersymmetry and $G =\mathbb{R}^+$ duality symmetry and the
IIB theory with (2,0) supersymmetry and $G = S L (2, \mathbb{R})$ duality
symmetry. The IIA theory is obtained by reducing 11-dimensional supergravity
on a circle, with the $G =\mathbb{R}^+$ duality symmetry emerging form a
combination of internal diffeomorphisms and the trombone symmetry. However,
there is only one maximal supergravity in $D = 9$ so the same theory is
obtained by reducing IIA or IIB supergravity on a circle, or from reducing
11-dimensional supergravity on $T^2 .$ The 9-dimensional supergravity action
has duality symmetry $S L (2, \mathbb{R}) \times \mathbb{R}^+$. This can be
understood from reducing 11-dimensional supergravity on $T^2$ to give a $S L
(2, \mathbb{R}) \times \mathbb{R}^+$ symmetry, while reducing the IIB theory
on $S^1$ gives the $G = S L (2, \mathbb{R})$ 10-dimensional duality combined
with a $\mathbb{R}^+$ symmetry associated with diffeomorphisms on $S^1$.

The two $D = 10$ supergravity theories are the low energy effective field
theories of the IIA and IIB superstring theories and these 2 theories, when
compactified on a circle, are T-dual: the IIA string theory compactified on a
circle of radius $R$ physically the same as the IIB string compactified on a
circle of radius $1 / R$, with the Kaluza-Klein modes of one theory
represented as winding modes of the dual theory. For the supergravity limit,
the winding modes are absence and so the duality only applies to the sector
without momentum on the internal circle, i.e. to the massless sector in 9
dimensions, which is governed by the 9-dimensional supergravity.

\subsection{8 dimensions}

From (\ref{glnr}), compactifying 11-dimensional supergravity on $T^3$ gives a
theory with
\begin{equation}
  \mathbb{R}^+ \times S L (3, \mathbb{R}) 
\end{equation}
symmetry. The same $d = 8$ theory can be obtained by reducing the IIB theory
on $T^2$, giving a symmetry
\begin{equation}
  S L (2, \mathbb{R})  \times S L (2, \mathbb{R}) 
\end{equation}
with one $S L (2, \mathbb{R})$ from the 10-dimensional duality and $S L (2,
\mathbb{R}) \times \mathbb{R}^+$ from the reduction on $T^2 .$

As a result, the $d = 8$ theory should have both $S L (3, \mathbb{R})$ and $S
L (2, \mathbb{R})  \times S L (2, \mathbb{R}) $ symmetry, and as these do not
commute, the full symmetry group $G$ must be larger and contain the group
generated by these two subgroups. The fields (other than the scalars) should
then be in representations of $G$. The supergravity has three 2-form gauge
fields transforming as a {\textbf{3}} under $S L (3, \mathbb{R})$ and as a
({\textbf{2,1}})+({\textbf{1,1}}) under $S L (2, \mathbb{R})  \times S L
(2, \mathbb{R})$. It has six vector fields transforming as a {\textbf{3+3}}
under $S L (3, \mathbb{R})$ and as a ({\textbf{2,2}})+({\textbf{1,2}})
under $S L (2, \mathbb{R})  \times S L (2, \mathbb{R})$. There are 8 scalar
fields which should take values in a coset space $G / H$.

The group generated by the $S L (3, \mathbb{R})$ and $S L (2, \mathbb{R}) 
\times S L (2, \mathbb{R}) $transformations is
\[ G = S L (3, \mathbb{R})  \times S L (2, \mathbb{R}) \]
with the scalars taking values in
\begin{equation}
  \frac{G}{H} = \frac{S L (3, \mathbb{R})}{S O (3)} \times \frac{S L (2,
  \mathbb{R})}{U (1)}
\end{equation}
The 2-form gauge fields transform as a ({\textbf{3,1}}) under $S L (3,
\mathbb{R})  \times S L (2, \mathbb{R}) $ while the vector fields transform as
a ({\textbf{3,2}}) under $S L (3, \mathbb{R})  \times S L (2, \mathbb{R})$.
The fermions transform under the double cover $\tilde{H} = {Spin} (3)
\times U (1)$ of $H = S O (3) \times U (1)$ (with the $U (1) $ in $\tilde{H}$
the double cover of the $U (1)$ in $H$).

\subsection{7 dimensions}

Reducing from 11 to 7 dimensions gives a duality
\[ S L (4, \mathbb{R})
 \times \mathbb{R}^+
 \]
while reducing from 8-dimensional supergravity gives a symmetry
\[ S L (3, \mathbb{R}) \times S L (2, \mathbb{R}) \times \mathbb{R} \]
The  reduction from $d=8$ gives five 2-form gauge fields in the
({\textbf{3,1}})+({\textbf{1,2}}) of $S L (3, \mathbb{R}) \times S L (2,
\mathbb{R})$ while the reduction of 11-dimensional supergravity gives four
2-forms in the {\textbf{4}} of $S L (4, \mathbb{R})$ plus a 3-form gauge
field. This 3-form can be dualised to a 2-form while maintaining the $S L (4,
\mathbb{R})$ symmetry to give five 2-form gauge fields in the {\textbf{4+1}} of $S
L (4, \mathbb{R})$.

The formulation of the theory with five 2-forms should have both $S L (3,
\mathbb{R}) \times S L (2, \mathbb{R})$ and $S L (4, \mathbb{R}) $ symmetry
and these close on the duality symmetry
\[ G = S L (5, \mathbb{R}) \]
with the scalars taking values in the coset
\begin{equation}
  \frac{G}{H} = \frac{S L (5, \mathbb{R})}{S O (5)}
\end{equation}
 The 2-forms
transform in the {\textbf{5}} of $S L (5, \mathbb{R})$.

\section{Reduction to $d\leq 2$ dimensions: the Geroch group and  extensions}

Consider pure gravity reduced from $D=4$ dimensions (with coordinates $(x^\mu,y^m)$ with $m=1,2$) to $d=2$.  
From the previous discussion, this theory should have $GL(2,\mathbb{R})$ symmetry.
Reducing first on $y^2$ gives a 3-dimensional theory with Ehlers symmetry $G_2=SL(2,\mathbb{R})$ and this will still be a symmetry after the further reduction on $y^1$. Similarly reducing first on $y^1$ gives a 3-dimensional theory with Ehlers symmetry $G_1=SL(2,\mathbb{R})$ and this will still be a symmetry after the further reduction on $y^2$. Thus the symmetry in $d=2$ will include the two Ehlers symmetries $G_1,G_2$
together with the \lq geometric'  $GL(2,\mathbb{R})$ symmetry. These do not commute and together generate an infinite dimensional group, known as the Geroch group \cite{Geroch:1970nt}. This is in fact an affine  extension of $SL(2,\mathbb{R})$ to an affine $SL(2,\mathbb{R})$ Kac-Moody symmetry.

Similarly, for maximal supergravity compactified from $D=4$ dimensions to $d=2$, there is a  \lq geometric'  $GL(2,\mathbb{R})$ symmetry together with
$G_2=E_8$ from first reducing on $y^2$ and $G_1=E_8$ from first reducing on $y^1$. These generate an infinite group $E_9$ which is an affine extension of $E_8$ to an affine $E_8$ Kac-Moody symmetry.

This argument can be continued further. 
Compactifying pure gravity  from $D=4$ to $d=1$ gives a \lq geometric'  $GL(3,\mathbb{R})$ symmetry together with three  Geroch groups, $G_1$ from compactifying first on $y^2,y^3$,  $G_2$ from compactifying first on $y^1,y^3$ and $G_3$ from compactifying first on $y^2,y^1$. The three Geroch groups together with $GL(3,\mathbb{R})$ then generate a duality group.
Similarly, for maximal supergravity reduced from $D=4$ to $d=1$ gives a \lq geometric'  $GL(3,\mathbb{R})$ symmetry together with three  $E_9$ groups, $G_1,G_2,G_3$ and these together generate a group $E_{10}$. See \cite{Damour:2002cu} for further discussion of $E_{10}$ symmetry.
Finally, for maximal supergravity reduced from $D=4$ to $d=0$ gives a \lq geometric'  $GL(4,\mathbb{R})$ symmetry together with four  $E_{10}$ groups, $G_1,G_2,G_3$ and these together generate a group $E_{11}$. See \cite{West:2003fc} for further discussion of $E_{11}$ symmetry and its application to the 11-dimensional theory.

\section{General Compactifications and $K 3$ Compactifications}

\subsection{Compactifiying on a general manifold}

As has been reviewed in previous sections, compactifying supergravities with
16 or 32 supersymmetries on $T^n$ gives a supergravity with a duality symmetry
group $G$ and scalars taking values in a coset space $G / H$. In a general
compactification, $T^n$ is replaced by an $n$-dimensional manifold
$\mathcal{M}_n$. The group of large diffeomorphisms for $T^n$ is $G L (n,
\mathbb{Z})$ and this gives rise to a duality symmetry $G L (n, \mathbb{Z})
\subseteq G$ for toroidal compactifications. For $\mathcal{M}_n$, the group of
large diffeomorphisms $\textrm{Diff}_0 (\mathcal{M}_n)$ will be a different
(possibly trivial) group, so for compactification on $\mathcal{M}_n$ any
duality symmetry is expected to include $\textrm{Diff}_0 (\mathcal{M}_n)$ but
not $G L (n, \mathbb{Z})$ in general. If there is a $p$-form gauge field $A$
appearing only through the field strength $F = d A$, for the compactification
on $T^n$ there is a $\mathbb{R}^q$ symmetry with $q = \left(\begin{array}{c}
  n\\
  p
\end{array}\right)$ under which $A$ is shifted by a $p$-form with constant
components.

For compactification on $\mathcal{M}_n$, there is a symmetry under
\begin{equation}
  \delta A = \Omega
\end{equation}
where $\Omega$ is a closed form, $d \Omega$=0. Transformations in which
$\Omega$ is exact, $\delta A = d \Lambda$, are gauge transformations but
transformations in which $\Omega$ is closed but not exact can be thought of as
large gauge transformations. The cohomology classes $H^p (\mathcal{M}_n,
\mathbb{R})$ are represented by harmonic $p$-forms. Choosing a basis $\{
\omega_i \}$ for the harmonic $p$-forms on $\mathcal{M}_n$ with $i = 1,
\ldots, b_p$ where $b_p$ is the $p$'th Betti number, the large gauge
transformations can be written as
\begin{equation}
  \delta A = \lambda^i \omega_i
\end{equation}
with constant parameters $\lambda^i .$ Then there is an $\mathbb{R}^{b_p}$
symmetry. Thus the $G L (n, \mathbb{Z}) \ltimes \mathbb{R}^q$ symmetry of the
$T^n$ reduction becomes a $\textrm{Diff}_0 (\mathcal{M}_n) \ltimes
\mathbb{R}^{b_p}$ symmetry for compactification on $\mathcal{M}_n$. Thus the
duality symmetry for supergravity compactification on $\mathcal{M}_n$ (if any)
will not be the same as that for compactification on $T^n$ in general. For
more general theories in which $F = d A + \cdots$ with Chern-Simons
corrections, the story is a little more complicated and non-abelian groups can
emerge; see \cite{Cremmer:1997ct,Cremmer:1998px} for further discussion.

The target space of the scalar sigma model is a moduli space for the metric
and gauge fields on the internal manifold. For torus reductions, $G L (n,
\mathbb{R}) / O (n)$ is the moduli space of constant metrics on $T^n$, $O (n,
n) / O (n) \times O (n)$ is the moduli space of constant metrics and constant
$B$-fields on $T^n$ while the moduli space of constant metrics and constant
$3$-form gauge fields on $T^n$ contributes to the coset spaces $E_n / K_n$
(with $K_n$ the maximal compact subgroup of $E_n$) discussed in earlier
sections. For general $\mathcal{M}_n$, the correspoding moduli spaces for
metrics and $p$-form gauge fields is not a coset space and so the structure is
rather different.

Compactification on $T^n$ preserves all of the supersymmetry of the original
supergravity theory. For theories with at least 16 supersymmetries, the
supergravity theory is essentially unique. For the corresponding toroidal
string theory compactification, the low energy effective supergravity theory
is unique and so the form of the moduli space $G / H$ is preserved in the
quantum theory. For compactification on $\mathcal{M}_n$, some or all of the
supersymmetry is broken in general. If the compactified \ theory has less than
16 supersymmetries, then the low energy effective theory will receive quantum
corrections in general, and in particular the quantum moduli space will be
different form the classical one.

\subsection{$K 3$ compactification}

Consider 11-dimensional supergravity compactified to $d = 7 - k$ dimensions on
$K 3 \times T^k$, first considered for $k=3$ in \cite{Duff:1983vj}. This preserves half the supersymmetry as $K 3$ has holonomy
$S U (2)$ and so has 2 covariantly constrant spinors. The resulting
supergravity theory has 16 supersymmetries and so must consist of half-maximal
supergravity in $d$ dimensions coupled to some number $r$ of vector
multiplets. As $K 3$ has no isometries, the gauge group is abelian. For $d >
4,$ the resulting theory then has duality group
\begin{equation}
  G = S O (n, n + r) \times \mathbb{R}^+
\end{equation}
with $n = 10 - d$ and with $n (n + r) + 1$ scalars in the coset
\begin{equation}
  \frac{G}{H} = \frac{S O (n, n + r)}{S O (n) \times S O (n + r)} \times
  \mathbb{R}^+
\end{equation}
and with $2 n + r$ vector fields.

Consider first the compactification to $d = 7$ so that $n = 3$ and there are
$6 + r$ vector fields and $3 (3 + r) + 1$ scalar fields. $K 3$ has trivial
$H^1$ and $H^3$, but has $H^2 =\mathbb{R}^{22},$ and a basis $\omega_i$ is
given by the 3 self-dual and 19 anti-self-dual harmonic 2-forms. Then the
3-form $A$ gives rise to vector fields 22 vector fields $B^i$ through the
ansatz
\begin{equation}
  A = \omega_i \wedge B^i
\end{equation}
No massless vector fields arise from the metric, as $K 3$ has no isometries.
Thus $r = 16$ and the number of scalars is 58. This is precisely the dimension
of the moduli space of metrics on $K 3$. \ The factor $\mathbb{R}^+$ is the
modulus given by the volume of the $K 3$ so that the moduli space of metrics
of fixed volume is
\begin{equation}
  \frac{S O (3, 19)}{S O (3) \times S O (19)}
\end{equation}
Remarkably, supergravity predicts that the $K 3$ moduli space is this coset
space.

For $n = 4, d = 6$ similar arguments give
\begin{equation}
  \frac{S O (4, 20)}{S O (4) \times S O (20)}
\end{equation}
as the moduli space of a metric and $B$-field on $K 3$.
Thus compactification on $K 3 \times T^k$ to $d = 10 - n > 4$ dimensions gives
a theory with duality group
\begin{equation}
  G = S O (n, n + 16) \times \mathbb{R}^+
\end{equation}
and scalar coset
\begin{equation}
  \frac{G}{H} = \frac{S O (n, n + 16)}{S O (n) \times S O (n + 16)} \times
  \mathbb{R}^+
\end{equation}
For compactification on $K 3 \times T^3$ to $d = 4$ this is enhanced (after
dualising the $B$-field) to
\begin{equation}
  G = S O (6, 22) \times S L (2, \mathbb{R})
\end{equation}
with scalar coset
\begin{equation}
  \frac{G}{H} = \frac{S O (6, 22)}{S O (6) \times S O (22)} \times \frac{S L
  (2, \mathbb{R})}{U (1)}
\end{equation}

\section{Timelike Reductions} \label{timelike}

Instead of compatifying on $n$spatial directions, one can formally reduce on
time and $n - 1$ spatial directions, corresponding to reducing on a Lorentzian
signature torus $T^{n - 1, 1}$ 
(or on $\mathbb{R})^{n - 1, 1}$)
to a Riemannian signature theory with $d = D -
n$ spatial dimensions. Such compactifications were investigated in \cite{Hull:1998br} and in
each case, the duality group $G$ remains the same as in the reduction on $T^n$
but the subgroup $H$ changes to a different real form of the same complex
group resulting in a coset space $G / H$ which is non-compact. In each case,
$H$ is non-compact and the R-symmetry group, which is a double cover
$\tilde{H}$ of $H$ is also non-compact.

Compactifying gravity on $T^n$ gave the duality group $G / H = G L (n,
\mathbb{R})^+ / S O (n)$ while the timelike reduction gives
\begin{equation}
  \frac{G}{H} = \frac{G L (n, \mathbb{R})^+}{S O (n - 1, 1)} \label{ghnrt}
\end{equation}
with the rotation group $S O (n)$ replaced with the Lorentz group $S O (n - 1,
1)$. For compactification from 4 to 3 dimensions, the Ehlers symmetry remains
$S L (2, \mathbb{R})$ but the coset space becomes $S L (2, \mathbb{R}) / S O
(1, 1)$.

For compactification of 11-dimensional supergravity on $T^{n - 1, 1}$ the
duality group $G$ remains $E_n$ but the R-symmetry groups are as given in  table 3.
For the half-maximal supergravities (\ref{sonmaa}) is replaced with \cite{Hull:1998br}
\begin{equation}
  \frac{S O (n, m)}{S O (n - 1, 1) \times S O (m - 1, 1)} \times \mathbb{R}^+
  \label{sonm2}
\end{equation}

\begin{table}
\[ \begin{array}{|c|c|c|c|}
     \hline
     d & G & H & \tilde{H}\\
     \hline
     10 & \mathbb{R}^+ & 1 & 1\\
     \hline
     9 & S L (2, \mathbb{R}) \times \mathbb{R}^+ & S O (1, 1) & S p i n (1,
     1)\\
     \hline
     8 & S L (3, \mathbb{R}) \times S L (2, \mathbb{R}) & S O (2, 1) \times S
     O (2) & S p i n (2, 1) \times S p i n (2)\\
     \hline
     7 & S L (5, \mathbb{R}) & S O (3, 2) & S p i n (3, 2)\\
     \hline
     6 & S p i n (5, 5) & S O (5, \mathbb{C}) & S p i n (5, \mathbb{C})\\
     \hline
     5 & E_6 & U S p (4, 4) /\mathbb{Z}_2 & U S p (4, 4)\\
     \hline
     4 & E_7 & S U^{\ast} (8) /\mathbb{Z}_2 & S U^{\ast} (8)\\
     \hline
     3 & E_8 & S p i n^{\ast} (16) /\mathbb{Z}_2 & S p i n^{\ast} (16)\\
     \hline
   \end{array} \]
   		\caption{ Duality groups for 11-dimensional supergravity compactified on the Lorentzian torus $T^{n-1,1}$}
\end{table}


\section{Non-Perturbative String Theory and U-Duality}

\subsection{The duality revolution}

Superstring theory was introduced as a perturbation theory in a dimensionless
coupling constant $g$ as a theory of fundamental superstrings propagating in
10-dimensional spacetime. The tension of the strings is set by a dimensionful
parameter $T$. The low energy effective field theory governing the massless
degrees of freedom is given by a supergravity theory in $D = 10$ dimensions.
For the IIA string it is (1,1) supergravity, for the IIB string it is (2,0)
supergravity and for the heterotic and type I strings it is (1,0) supergravity
coupled to (1,0) super-Yang-Mills theory. From 1994, huge progress was made in
understanding the non-perturbative structure of superstring theories in what
has been called the 2nd superstring revolution. For a history of the duality revolution and the work that led up to it,  see \cite{HullTalk}, while for 
a comprehensive review and an extensive set of references, see \cite{Obers:1998fb}.

In this work, it is assumed that there is a non-perturbative completion of
superstring theory and evidence is sought about its structure. In particular,
non-perturbative symmetries are sought and these are then used to infer
information about the spectrum. In the absence of a fundamental formulation of
the non-perturbative theory, much of this development is conjectural but by
now a considerable amount of evidence has been amassed in support of these
conjectures.

For any of these string theories compactified on $T^q$, the result is a theory
with vacuum $\mathbb{R}^{9 - q, 1} \times T^q$ whose low-energy effective
action is a supergravity in $d = 10 - q$ dimensions with 16 or 32
supersymmetries and so is one of the theories discussed in earlier sections (with $q=n-1$).
The aim is to find symmetries of the non-perturbative theory with vacuum
$\mathbb{R}^{9 - q, 1} \times T^q$, beyond the expected diffeomorphism
invariance, local supersymmetry and gauge symmetry.

The key idea is to argue that there are certain properties that are protected
by supersymmetry and so can be extrapolated from weak to strong coupling 
\cite{Hull:1994ys,Sen:1993zi,Sen:1993sx,Sen:1992fr}.

First, as the supergravity theory is unique, it cannot be changed by quantum
corrections and so any symmetry of the full theory must be a symmetry of the
supergravity theory. Thus the symmetry group must be contained in the duality
group $G$ and local $\tilde{H}$ R-symmetry. Secondly, the BPS spectrum can be
extrapolated from weak coupling to all values of $g$. A BPS state is a state
invariant under some supersymmetries and so as a result its mass saturates a
BPS bound given in terms of the charges it carries. As it has the minimum mass
for a given set of charges, it must be stable and, if it preserves sufficient
supersymmetry, it does not receive quantum corrections. As a result, the
spectrum of BPS states can be extrapolated from weak to strong coupling, so
that a BPS state identified in the weak coupling theory persists at all values
of the coupling. The charges carried by the BPS states fit into
representations of the duality group $G$ as they are charges coupling to the
supergravity gauge fields which themselves fit into representations of $G$.
Next, these charges obey Dirac quantisation conditions in the quantum theory
and so take values in a discrete lattice. This restricts the symmetry group
$G$ to be the discrete subgroup of $G$, denoted $G (\mathbb{Z})$, 
that preserves the charge lattice.
We will now examine how this
works.

\subsection{Perturbative symmetry: T-duality}

In each case, $G$ has a subgroup $G L (q, \mathbb{R})$ as has been seen in
section \ref{PureGrav}. If the internal space is taken to be $\mathbb{R}^q$
then these act as diffeomorphisms on $\mathbb{R}^q$ (combined with the
trombone symmetry) but for an internal space that is $T^q$ this is broken to
the discrete subgroup $G L (q, \mathbb{Z})$ which is the subgroup that acts as
diffeomorphosms of $T^q .$ Thus although the $d$-dimensional supergravity has
symmetry $G$, the full string theory has only the discrete subgroup $G
(\mathbb{Z}) .$

The relevant BPS states here are the Kaluza-Klein modes. Each massless field
$\Psi (x^{\mu}, y^m)$ in 10-dimensions has a Fourier expansion
\begin{equation}
  \Psi (x, y) = \sum_{N_m} \psi_{_{N_1 N_2 \ldots .N_q}} (x) \exp (i y^m N_m)
\end{equation}
to give an infinite set of fields $\psi_{_{N_1 N_2 \ldots .N_q}}$ labelled by
integers $N_1, N_2, \ldots .N_q$. For the simple case of a rectangular torus
with diagonal internal metric
\begin{equation}
  h_{m n} = {\rm{diag} }(R_1^2, R_{2 }^2, \ldots, R_q^2)
\end{equation}
(so that the radius of the $y^m$ circle is $R_m$) the momentum of $\psi_{_{N_1
N_2 \ldots .N_q}}$ in the $y^m$ direction is quantised in units of the inverse
radius and given by
\begin{equation}
  p_m = \frac{N_m}{R_m}
\end{equation}
The mass of $\psi_{_{N_1 N_2 \ldots .N_q}}$ is then
\begin{equation}
  M_{_{N_1 N_2 \ldots .N_n}} = \sqrt{p_{1 }^2 + p_2^2 + \cdots + p_{q }^2}
\end{equation}
For a general torus, the mass is
\begin{equation}
  M_{_{N_1 N_2 \ldots .N_n}} = \sqrt{h^{m n} N_m N_n} \label{kkbps}
\end{equation}
The $\psi_{_{N_1 N_2 \ldots .N_q}}$ are the Kaluza-Klein modes and
(\ref{kkbps}) is the BPS mass condition that they satisfy. The momenta $p_m$
are the BPS charges. The massless fields in $10$ dimensions for type II
strings are in the supergravity multiplet and as a result the BPS states fit
into ultra-short massive supergravity multiplets (consisting of a massive
spin-two field together with massive fields of lower spin). For the type I and
heterotic theories, they fit into (1,0) massive supergravity and vector
multiplets.

If the internal space were $\mathbb{R}^q$, the momenta $(p_1, p_2, \ldots
p_q)$ would be continuous and would transform under the $G L (q, \mathbb{R})$
diffeomorphisms in the vector representation $\mathbf{q}$. For the internal
space $T^q$, the momenta take values in a discrete lattice and as a result $G
L (q, \mathbb{R})$ is broken to the discrete subgroup preserving the lattice.
Regarding this lattice as the set $\mathbb{Z}^q$ of points $(N_1, \ldots,
N_n)$, the subgroup of $G L (q, \mathbb{R})$ of lattice automorphisms is the
infinite discrete group $G L (q, \mathbb{Z}) $ of invertible matrices with
integer components whose inverse is also integer-valued.

Next, the theory of gravity plus a 2-form gauge field and dilaton compactified
on $T^q$ has a symmetry $S O (q, q) \times \mathbb{R}^+$ and this in fact
extends to a symmetry of the whole supergravity theory for each superstring,
so that in each case $S O (q, q) \times \mathbb{R}^+$ is a subgroup of $G$. It
is known that the discrete subgroup $S O (q, q ; \mathbb{Z})$ is an exact
perturbative symmetry of toroidally compactified bosonic heterotic  theory, known as
T-duality symmetry; see \cite{Giveon:1994fu} for a review. 
For the heterotic or type I strings
(at a point in the moduli space at which the unbroken gauge group is abelian),
this is extended to $S O (q, q + 16 ; \mathbb{Z})$ \cite{Narain:1985jj,Narain:1986am}, while for type II strings the
$Spin(q,q)$ subgroup of $E_{q+1}$ is broken to the 
$Spin(q,q ;\mathbb{Z})$ type II T-duality group \cite{Hull:1994ys}.

The relevant BPS states in this case are the Kaluza-Klein modes and the string
winding modes. A string winding $\tilde{N}^m$ times round the $y^m$ circle of
radius $R_m$ and has mass $w_m = 2 \pi T \tilde{N}^m R_m$. If it winds round
all of the circles its mass (for the rectangular torus) is
\begin{equation}
  M = \sqrt{w_{1 }^2 + w_2^2 + \cdots + w_{q }^2}
\end{equation}
For a general state with both momentum and winding on a general torus with
moduli $h_{m n}, B_{m n}$, the mass is \cite{Giveon:1994fu}
\begin{equation}
  M^2 =\mathcal{K}^t \mathcal{M}\mathcal{K}
\end{equation}
where $\mathcal{M}$ is the matrix (\ref{miss}) and $\mathcal{K}$ is the
2n-vector of integers
\begin{equation}
  \mathcal{K}= \left(\begin{array}{c}
    N_m\\
    \tilde{N}^m
  \end{array}\right)
\end{equation}
If $\mathcal{K}$ were a vector of $2q$ real numbers, this would be invariant
under $O (q, q)$ transformations (\ref{mtrans}) with
\begin{equation}
  \mathcal{K} \rightarrow \Lambda^{- 1} \mathcal{K}
\end{equation}
However, as $\mathcal{K}$ takes values in the lattice $\mathbb{Z}^{2 q}$, this
invariance is broken to the subgroup $O (q, q ; \mathbb{Z})$ of invertible $O
(q, q)$ matrices satisfying (\ref{onncon}) with integer components whose
inverse is also integer-valued. 
In addition to these perturbative BPS states, there are non-perturbative ones carrying RR charge; see below.
These RR BPS states in fact fit into spinor representations of the T-duality group, so that it is in fact $Spin(q,q ;\mathbb{Z})$ for type II strings.

The Kaluza-Klein modes and the winding modes are perturbative BPS states and
the T-duality has been proven to be a symmetry of perturbative string theory; see
\cite{Giveon:1994fu} and references therein. 
The fact that these BPS states of the weakly coupled theory can be
extrapolated to BPS states for all values of the coupling supports the
conjecture that $Spin (q, q ; \mathbb{Z})$ is in fact a symmetry of the full
non-perturbative theory. Certainly, the $G L (q, \mathbb{Z})$ subgroup of this
that corresponds to invariance under large diffeomorphisms of the torus should
be a symmetry of the full non-perturbative theory.

\subsection{S-duality}

We turn now to symmetries that mix perturbative and non-perturbative states
and so are not symmetries of the perturbative theory. The first case that was
considered was the heterotic string compactified on $T^6$ to 4 dimensions.
The low energy effective field theory is $\mathcal{N}= 4$ supergravity coupled
to $\mathcal{N}= 4$ super-Yang-Mills. For configurations in which the gauge
group is spontaneously broken to an abelian subgroup $U (1)^{28}$, the supergravity  duality
group is
\begin{equation}
  G = S O (6, 22) \times S L (2, \mathbb{R})
\end{equation}
As has been seen, the $S O (6, 22)$ is broken to the T-duality group $S O (6,
22 ; \mathbb{Z}) .$ We now turn to the classical $S L (2, \mathbb{R})$ symmetry, which had been conjectured to be broken to a $S L (2, \mathbb{Z})$ symmetry in the quantum string theory \cite{Font:1990gx,Schwarz:1993mg,Schwarz:1993vs}.

Consider first the $\mathcal{N}= 4$ super-Yang-Mills theory. Classical abelian
$\mathcal{N}= 4$ super-Yang-Mills has a classical $S L (2, \mathbb{R})$
symmetry of the equations of motion, acting as electromagnetic duality with
the pair $(\mathcal{F} _{\mu \nu}, \mathcal{G} ^{\mu \nu})$ (the field
strength $\mathcal{F}$ and its dual $\mathcal{G}$ given by (\ref{gdust}))
transforming as an $S L (2, \mathbb{R})$ doublet for each $U (1)$ factor. The
non-abelian theory with gauge symmetry spontaneously broken to an abelian
subgroup has BPS states consisting of W-bosons and magnetic monopoles \cite{Witten:1978mh}. As
had been pointed out by Osborne \cite{Osborn:1979tq}, in the  $\mathcal{N}= 4$ theory both the
W-bosons and the magnetic monopoles fit into massive vector supermultiplets,
so that this is a theory in which the electromagnetic duality conjectured by
Montonen and Olive \cite{Montonen:1977sn} could apply. The electric and magnetic charges transform
into each other under $S L (2, \mathbb{R})$ -- for each unbroken $U (1)$ the
electric and magnetic charges fit into a doublet of $S L (2, \mathbb{R})$. The
Dirac-Schwinger-Zwanziger quantisation condition on the electric charges
carried by dyons implies that the electric and magnetic charges must take
values in a lattice, so that the $S L (2, \mathbb{R})$ must be broken to at
most the discrete subgroup $S L (2, \mathbb{Z})$ preserving the lattice. In
\cite{Sen:1994yi}, Sen provided important non-trivial evidence that the spectrum of BPS dyons
is precisely in agreement with the predictions of $S L (2, \mathbb{Z})$
symmetry. This symmetry became known as S-duality and is non-perturbative as
it relates perturbative states such as the W-bosons to non-perturbative states
such as the magnetic monopoles. In the weakly coupled theory, the magnetic
monopoles are understood as solitons. They are solutions of the classical
theory that correspond to BPS quantum states.

In \cite{Font:1990gx,Schwarz:1993mg,Schwarz:1993vs}
 it was conjectured that $S L (2, \mathbb{Z})$ is an exact symmetry of 
the heterotic string compactified on $T^6$, so that the supergravity symmetry
$G = S O (6, 22) \times S L (2, \mathbb{R})$ is broken to the discrete
subgroup
\begin{equation}
\label{n4du}
  G (\mathbb{Z}) = S O (6, 22 ; \mathbb{Z}) \times S L (2, \mathbb{Z})
\end{equation}
Further evidence supporting this was given in \cite{Sen:1993zi, Sen:1993sx, Sen:1992fr}.

\subsection{U-duality}

My work with Townsend \cite{Hull:1994ys} started with the question of what the $\mathcal{N}=
8$  generalisation of the $\mathcal{N}=
4$   duality symmetry (\ref{n4du}) could be. Type II string theory
compactified on $T^6$ has a low energy effective field theory given by $d = 4,
\mathcal{N}= 8$ supergravity with duality group $G = E_7$. There are 28
abelian vector fields, so there are 28 electric charges and 28 magnetic
charges which fit into the {\textbf{56}} representation of $E_7$. These 56
charges are quantised so that they must lie in a lattice $\mathbb{Z}^{56}$.
Thus, if there are BPS states carrying each of the 56 charges, then the
duality group must be broken to at most the discrete subgroup of $E_7$ that is
an automorphism of the lattice.

This group was called $E_7 (\mathbb{Z})$ and can be understood as follows.
The group $E_7$ is the subgroup of $S p (56, \mathbb{R})$ preserving a certain
4-form. The group $S p (56, \mathbb{R})$ acts on its fundamental
representation through $56 \times 56$ matrices and the subgroup preserving the
lattice $\mathbb{Z}^{56}$ is $S p (56, \mathbb{Z})$, \ the group of matrices
in $S p (56, \mathbb{R})$ with integer entries whose inverses are also
matrices of integers. The group $E_7 (\mathbb{Z})$ is the subgroup of $S p
(56, \mathbb{Z})$ preserving the 4-form \cite{Hull:1994ys}. This group is certainly
non-trivial: the subgroup $S O (6, 6)$ of $E_7$ is broken to the T-duality
group $S O (6, 6 ; \mathbb{Z})$ while the $G L (7, \mathbb{R})$ subgroup is
broken to the group $G L (7, \mathbb{Z})$ of torus diffeomorphisms so that
$E_7 (\mathbb{Z})$ contains both $S O (6, 6 ; \mathbb{Z})$ and $G L (7,
\mathbb{Z})$.

To see whether this is the case requires understanding the BPS spectrum and in
particular whether all 56 charges are carried by BPS states. There are 12 of
the 56 charges which are carried by the perturbative string: the 6 string
momenta $p_m$ and 6 winding charges $w^m$. These are electric charges,
coupling to the vector fields $A_{\mu}^m, B_{\mu m}$ that arise in the NS-NS
sector of the string spectrum. The remaining charges are non-perturbative.
These consist of the magnetic charges for the NS-NS vectors fields $A_{\mu}^m,
B_{\mu m} $together with the electric and magnetic charges for the 16 vector
fields $C_{\mu}^i$ that arise in the RR sector of the string spectrum. The
perturbative string has no states that carry the RR electric charges so both
the electric and magnetic RR charges could only be carried by non-perturbative
states.

Gibbons \cite{Gibbons:1981ux} had argued that extreme black holes could be regarded as solitons
of gravity and  supergravity as they are stable and have zero temperature. In
\cite{Gibbons:1982fy} it was shown that there is a BPS bound for supergravity and that extreme
black holes saturating the bound are stable, so that they can be regarded as
BPS solitons. The $\mathcal{N}= 8$ supergravity has black holes carrying each
of the 56 charges and so there is a half-supersymmetric BPS soliton carrying
each of these charges. It was argued in \cite{Hull:1994ys} that the BPS solitons of the low
energy effective supergravity theory can be associated with BPS states of the
full string theory and that their charges should take values in the lattice
$\mathbb{Z}^{56}$. Thus the duality group should indeed be $E_7 (\mathbb{Z})$,
which was called U-duality \cite{Hull:1994ys}.

As discussed above, 12 of the electric charges are carried by perturbative
string states and the solution gives the supergravity field configuration
outside the fundamental string state. The black hole solutions carrying
magnetic charges for the vector fields $A_{\mu}^m, B_{\mu m}$ lift to
Kaluza-Klein monopole and NS 5-brane solutions of 10-dimensional supergravity,
and these are solitons of the 10-dimensional supergravity. These are genuine
solitons of the theory and can be associated with BPS states of the full
theory.

The black holes carrying each of the remaining 32 charges for the 16 RR
fields $C_{\mu}^i$ also lift to brane solutions in 10-dimensions. These were
subsequently shown by Polchinski \cite{Polchinski:1995mt} to be D-branes so that they correspond to 
boundary states for the perturbative string theory.

There was a similar story in each dimension \cite{Hull:1994ys}. In $d = 11 - n$ dimensions,
the supergravity solitons fit into representations of the duality group $G =
E_n$ and charge quantisation implies that this is broken to a discrete
U-duality group $E_n (\mathbb{Z})$ in the quantum theory. In particular, the
$S L (2, \mathbb{R})$ symmetry of the IIB supergravity becomes an $S L (2,
\mathbb{Z})$ symmetry of the full IIB string theory. (This was referred to as
a U-duality in \cite{Hull:1994ys}, but subsequently it is more often called an S-duality.)

As well as black hole solitons, there are black brane supergravity solitons.
An important consequence of the U-duality conjecture is that these must
correspond to BPS states of the full string theory. As seen above, for string
theory compactified on $T^6$, the solitons carrying the NS-NS electric charges
are related by U-duality to ones carrying NS-NS magnetic charges and ones RR
charges. Solitons carrying NS-NS electric charges are associated with 
perturbative string states carrying torus momentum $p_m$ or winding charge
$w^m$, and these are extrapolated to BPS states for all values of the string
coupling, even though the string picture is only reliable at weak coupling.
Then the U-dual supergravity solutions of wrapped branes should also
correspond to BPS brane states of the non-perturbative theory. Thus
non-perturbative string theory must be a theory of branes as well as strings,
and the U-duality between wound strings and wrapped branes means that they
should all be regarded as being on the same footing.

\section{String-String Dualities and 11 Dimensions} \label{String-String}

\subsection{String-String Dualities} 

There are important dualities between different string theories. For string
theory A compactified on a manifold $\mathcal{N}_A$ to be the same theory as
string theory B compactified on a manifold $\mathcal{N}_B$, there are a number
of necessary conditions that must be met, by similar arguments to those in the
previous section. If both compactifications preserve at least 16
supersymmetries, then the low energy effective supergravity actions should be
unique. Then the supergravity limit of A compactified on $\mathcal{N}_A$
should give the {\em{same}} supergravity theory as the supergravtity limit
of B compactified on $\mathcal{N}_B$. Second, the BPS spectra for the two
compactifications should be the same.

The first example is the IIA string theory compactified on a circle of radius
$R$ and the IIB string theory compactified on a circle of radius $R' = 1 / 2
\pi T R$. The supergravity limits of the two string theories are the (1,1) and
(2,0) supergravities in 10 dimensions, and the circle reductions of both give
the {\em{same}} 9-dimensional supergravity, so the first test is met. The
perturbative BPS spectrum of the IIA theory consists of momentum modes with
mass $N / R$ and winding modes with mass $2 \pi T \tilde{N} R$ for integers
$N, \tilde{N}$ while for the IIB string they consist of \ momentum modes with
mass $N' / R'$ and winding modes with mass $2 \pi T \widetilde{N'} R'$ for
integers $N', \widetilde{N'}$. The spectra are the same if one identifies
\begin{equation}
  R' = \frac{1}{2 \pi T R}, \quad N' = \tilde{N}, \quad \tilde{N}' = N
\end{equation}
so that the winding modes of the IIA string are identified with the momentum
modes of the IIB string and the winding modes of the IIB string are identified
with the momentum modes of the IIA string. Thus the {\em{perturbative}} BPS
spectra agree and it can be shown that the two perturbative string theories
are T-dual \cite{Dai:1989ua}: the IIA string compactified on a circle of radius R and the IIB
string theory compactified on a circle of radius $R' = 1 / 2 \pi T R$ are the
{\em{same}} string theory, with the momentum modes and winding modes
interchanged by the duality. Moreover, the non-perturbative spectra of BPS
branes is the same, supporting the conjecture that the duality extends to the
non-perturbative theories \cite{Hull:1994ys}. In the absence of a fundamental formulation of
either non-perturbative theory, this cannot be more than a conjecture, but
assuming the duality gives much information about the non-perturbative
structure of the two theories. There is a simlar T-duality between the two
heterotic string theories \cite{Dai:1989ua}: \ the $S O (32)$ heterotic string compactified
on a circle of radius $R$ is the same string theory as the $E_8 \times E_8$
heterotic string compactified on a circle of radius $R' = 1 / 2 \pi T R$.

Consider next the heterotic string compactified on $T^q$ for $q \geqslant 4$
(at a generic point in the moduli space at which the gauge symmetry is broken
to an abelian subgroup) and the IIA string compactified on $K 3 \times T^{q -
4}$. (Note that as the $S O (32)$ and $E_8 \times E_8$ heterotic strings are
dual on $S^1$, they are also dual on $T^4$ and so that both heterotic strings
give the same result here.) These are two very different string theories, but
the remarkable fact that they have the same supergravity effective action
suggests a surprising duality \cite{Hull:1994ys}. For both the heterotic string on $T^n$ and
the IIA string compactified on $K 3 \times T^{q - 4}$ the low-energy effective
action is the {\em{same}} supergravity theory coupled to abelian vector
multiplets with 16 supersymmetries, with duality group $G$ and scalar coset
space $G / H$ given by
\begin{equation}
  \frac{G}{H} = \frac{S O (q, q + 16)}{S O (q) \times S O (q + 16)} \times
  \mathbb{R}^+
\end{equation}
for $q < 6$, by  (\ref{cost})
for $q = 6$
and (\ref{cost8})
for $q = 7$. The two perturbative string theories are very different, so any
duality cannot be perturbative. In particular, the spectrum of perturbative
BPS states is different for the two theories. However, the full BPS spectrum
of the two theories is exactly the same, leading to the conjecture \cite{Hull:1994ys} that
the two theories are dual, defining the {\em{same}} non-perturbative
theory. It was later argued \cite{Witten:1995ex} that for $q = 4$ the heterotic theory on $T^4$
and the type IIA theory on $K 3$ are S-dual in the sense that the
strong-coupling limit of one is the weak-coupling limit of the other.

In this way, there are dualities linking the IIA and IIB strings, the $S O
(32)$ and $E_8 \times E_8$ heterotic strings, and the IIA and heterotic
strings. The remaining string theory is the type I string theory with gauge
group $S O (32)$. This has the same supergravity limit and BPS spectrum as the
$S O (32)$ heterotic string, supporting the conjecture that they could be
dual. It was argued in \cite{Witten:1995ex} that they are indeed  S-dual with the
strong-coupling limit of one giving the weak-coupling limit of the other. Thus
there are dualities linking all of the five superstring theories and as a
result they should be viewed as different perturbative limits of a unique
non-perturbative theory, which has become known as M-theory.

\subsection{11 dimensions and M-theory}

There had been considerable work on supermembranes in 11 dimensions as a possible generalisation of superstrings in 10 dimensions; see e.g. the account by Bergshoeff et al in this volume.
In \cite{Hull:1994ys}, it was pointed out that IIA supergravity compactified on $T^n$ or $K 3
\times T^r$ gives the same dimensionally reduced supergravity and the same BPS
brane spectrum as 11-dimensional supergravity compactified on $T^{n + 1}$ or $K
3 \times T^{r + 1}$, providing evidence for a duality between IIA string theory
and some theory whose low-energy effective action is 11-dimensional
supergravity. This was discussed further in \cite{Townsend:1995kk}. In \cite{Witten:1995ex} it was argued that the
strong coupling limit of IIA supergravity is an 11-dimensional theory whose
low-energy physics is governed by 11-dimensional supergravity. The BPS branes
of IIA string theory then arise from BPS branes of the 11-dimensional theory.
As a result, M-theory encompasses an 11-dimensional theory as well as the five
perturbative superstring theories, and from  \cite{Hull:1994ys}
this must be a theory of branes as well as strings.
This was the starting point for much
subsequent research.

\section{Times and Stars}

\subsection{The IIA* and IIB* Supergravities}

In section \ref{String-String} it was seen that cases in which two  supergravity compactifications  result in the same lower dimensional
supergravity is interesting and can be evidence for a new string duality. For
example, the circle reduction of the IIA (1,1) and IIB (2,0) supergravities
both give the same 9-dimensional supergravity, which provides evidence for the
T-duality between the corresponding IIA and IIB string theories. An important
role is played here by the uniqueness of 9-dimensional maximal supergravity.
The opposite of dimensional reduction from $D$ to $d$ dimensions has been
called {\em{oxidation}}: the lift of a $d$ dimensional theory to a $D > d$
dimensional one. Thus the 9 dimensional supergravity has two oxidations, one
with (1,1) local supersymmetry and one with (2,0) local supersymmetry.

Consider now the timelike reduction of the (1,1) and (2,0) supergravities. The
timelike reduction of IIA (1,1) supergravity to 9 Euclidean dimensions is the
same as the reduction of 11-dimensional supergravity on the Lorentzian torus
on $T^{1, 1}$ discussed in section \ref{timelike}. This gives a theory with duality $G = S
L (2, \mathbb{R}) \times \mathbb{R}^+$ and scalar coset
\begin{equation}
  \frac{G}{H} = \frac{S L (2, \mathbb{R})}{S O (1, 1)} \times \mathbb{R}^+
\end{equation}
The metric on $\frac{S L (2, \mathbb{R})}{S O (1, 1)}$ has Lorentzian
signature and is given by
\begin{equation}
  d s^2 = d \Phi^2 - e^{2 \Phi} d \chi^2
\end{equation}
so that the resulting scalar sigma model coupled to gravity has the lagrangian given by
\begin{equation}
  L = \sqrt{- g} \left( R - \frac{1}{2} (\partial \Phi)^2 + \frac{1}{2} e^{2
  \Phi} (\partial \chi)^2 \right) \label{dualact2}
\end{equation}
(instead of (\ref{dualact})) with the scalar $\chi$ having the wrong sign kinetic term. For the timelike
reduction from 10 dimensions, all fields are taken to be independent of the
time coordinate $t$, and the scalar field $\chi$ arises from the time
component $C_t$ of the vector field $C_M$ of IIA supergravity. However, the
timelike reduction of IIB supergravity gives theory with scalar coset
\begin{equation}
  \frac{G}{H} = \frac{S L (2, \mathbb{R})}{S O (2)} \times \mathbb{R}^+
\end{equation}
and is a {\em{different}} 9-dimensional Euclidean supergravity. Thus for
Euclidean signature, there are 2 9-dimensional maximal supergravities, which
can be called the 9A and the 9B supergravities. Thus the 9A theory has an
oxidation to (1,1) supergravity and the 9B theory has an oxidation to (2,0)
supergravity. 

However, if the 9A theory has a (2,0) oxidation, it 
 should be to a theory
with coset space $\frac{S L (2, \mathbb{R})}{S O (1, 1)}$ instead of the
$\frac{S L (2, \mathbb{R})}{S O (2)}$ of IIB supergravity so the oxidation
cannot be to
the usual IIB supergravity. If the 9B theory has a (1,1) oxidation, it
cannot be to  IIA supergravity as the kinetic term of the vector field in the 9B theory arising from  the time component $C_{\mu t}$ of the IIB RR 2-form in D=10 has the wrong sign in $d=9$ and so should lift to a vector field with a kinetic term of the wrong sign in $D=10$.
In \cite{Hull:1998vg} it was seen
that there are indeed new supergravity theories, the IIA* and IIB* theories,
with the IIB* supergravity the (2,0)-supersymmetric timelike oxidation of the
9A theory and the IIA* supergravity the (1,1)-supersymmetric timelike
oxidation of the 9B theory. In particular, the scalar coset space for the IIB*
supergravity is indeed $\frac{S L (2, \mathbb{R})}{S O (1, 1)}$. The starred
and unstarred theories are rather similar except for some sign changes in the
action. The bosonic spectra of the type II theories all contain the NS-NS
fields $g_{M N}, B_{M N}, \Phi$ together with RR $p$-form fields $C_p$ where $p
= 1, 3$ for the IIA,IIA* theories and $p = 0, 2, 4$ for the IIB,IIB* theories.
The bosonic action for the IIA* theory is similar to that of the IIA theory
but with the ``wrong'' sign for the RR fields $C_1, C_3 $ and the bosonic
action for the IIB* theory is similar to that of the IIB theory but with the
``wrong'' sign for the RR fields $C_0 = \chi, C_2, C_4$. 
It is remarkable that the IIA* and IIB* theories exist as supergravity theories, 
but the \lq wrong' signs in their kinetic terms means that they are
not unitary.

It was further argued that this lifted to a string theory T-duality \cite{Hull:1998vg}, with
the T-dual on a timelike circle of the IIA string theory giving a IIB* string
theory and the timelike T-dual the IIB string theory giving a IIA* string
theory. See  \cite{Hull:1998vg} for further discussion of the significance of these string theories.

\subsection{Lifting the IIA* theory to 11 dimensions: the 11* theory}

The IIA theory has an oxidation to 11-dimensional supergravity, so it is
natural to ask whether the IIA* theory also has an oxidation. There is indeed
such an oxidation, but it turns out to be a theory in a spacetime of signature
9+2 instead of the usual Lorentzian signature 10+1, so that the spacetime has
two time dimensions and 9 space ones. The bosonic action is of the form (\ref{11act}) but
the kinetic term for the 3-from gauge field has the ``wrong'' sign. The
gravitino in 10+1 dimensions is Majorana but this is not a consistent reality
condition in 9+2 dimensions. Instead, the  gravitino in 9+2 dimensions is
pseudo-Majorana \cite{Hull:1998ym}. This variant 11-dimensional supergravity can be referred
to as the 11* supergravity. In \cite{Hull:1998ym}, it was argued to be an effective action for
M*-theory, a regime of M-theory in spacetime of signature 9+2.

\subsection{Reducing the 11* theory: 10 dimensional supergravities with new
signatures}

The timelike reduction of 11* supergravity of course gives IIA* supergravity,
but the spacelike reduction gives a new supergravity in 8+2 dimensions, the
IIA$_{8 + 2}$ supergravity with a pseudo-Majorana gravitino.
Here and in what follows, in signature $s+t$, a supergravity will be referred to as a IIA$_{s+t}$ supergravity if the gravitino is a 32-component Majorana or pseudo-Majorana spinor and as a IIB$_{s+t}$ supergravity if the gravitino has 32 components of the same chirality, so that it consists of a pair of 16-component Majorana-Weyl spinors or  a 32-component symplectic Majorana-Weyl spinor. See \cite{Kugo:1982bn,Hull:1998ym} for definitions  and a discussion of which spinors are possible in which signatures.

 The spacelike reduction gives a theory in 7+2 dimensions which
in turn has a timelike oxidation to a IIB-like theory in 7+3 dimensions, the
IIB$_{7 + 3}$ theory. This has a symplectic
Majorana-Weyl gravitino. Then the spacelike reduction of this to 6+3
dimensions followed by a timelike IIA oxidation gives a IIA
theory in 6+4 dimensions, the IIA$_{6 + 4}$ theory. Then further iterations
of these constructions give 4 theories in signature 5+5, the IIA$_{5 +
5}$,IIB$_{5 + 5}$,IIA*$_{5 + 5}$,IIB*$_{5 + 5}$ theories. 9+1 and 5+5
(together with 1+9) are the only signatures in 10 dimensions that allow
Majorana-Weyl spinors. The actions for the IIA$_{5 + 5}$,IIB$_{5 +
5}$,IIA*$_{5 + 5}$,IIB*$_{5 + 5}$ supergravities are formally identical to the
IIA,IIB,IIA*,IIB* supergravities respectively, but with different spacetime
signatures. These theories were all found in \cite{Hull:1998ym}, where it was argued for each
there is a corresponding string theory, and that each of these
reduction-oxidation constructions lifts to a string theory T-duality. 

Note that not all candidate oxidations lead to supersymmetric theories.
For example, the 9-dimensional theory in 7+2 dimensions has a timelike oxidation to the IIB$_{7 + 3}$ theory and a spacelike oxidation to the
IIA$_{8 + 2}$ supergravity. However it does not have a spacelike oxidation to a IIB theory or a timelike oxidation to a IIA theory. One way of seeing this is that Weyl spinors are not possible in $8+2$ dimensions and Majorana or pseudo-Majorana spinors are not possible in 7+3 dimensions \cite{Kugo:1982bn,Hull:1998ym}.

\subsection{Oxidising the IIA$_{5 + 5}$ supergravity: the $11'$ theory.}

The IIA$_{5 + 5}$ theory has a spacelike oxidation to a theory in 6+5
dimensions, which can be referred to as the $11'$ theory. Further iterations
of the constructions above lead to 11-dimensional theories in signatures 5+6,
2+9 and 1+10, and for each 10-dimensional theory with signature $s + t$, there
is one with signature $t + s$. 
(See \cite{Duff:2006ix} for a discussion of signature reversal.) Remarkably, the theories that arise in this way
are the only possible supergravity theories. For example, in 11 dimensions, the
signatures 10+1,9+2 and 6+5 (plus their reverses) are the {\em{only}} ones
that have spinors with 32 real components. In other signatures the minimal
spinors have 64 real components and so supergravity is not possible.

The various 11-dimensional supergravities have a field content of a metric, a 3-
form gauge field  $A$ with field strength $F=dA$ 
and a gravitino which is Majorana for signatures $10 + 1,6 + 5,2 + 9$
and pseudo-Majorana for the mirror signatures $1 + 10,5 + 6,9 + 2$.
The bosonic
Lagrangian is of the form $R-F^2+\dots $ for signatures $10 + 1,6 + 5,2 + 9$
while the 3-form gauge field kinetic term has the wrong sign for the signatures
$1 + 10,5 + 6,9 + 2$, with lagrangian  $R+F^2+\dots $.

In \cite{Hull:1998ym} it was argued that all of these 10-dimensional supergravities are the
effective field theories for string theories, and that each of these
reduction-oxidation constructions arises from a string theory T-duality.
Moreover, there are further S-dualities \cite{Hull:1998ym}. The strong coupling limits of the
IIA theories were conjectured to give theories whose field theory limits are
the usual 11-dimensional supergravity together with the 11* and $11'$
theories; these theories were referred to as the M* and M$'$ theories \cite{Hull:1998ym}.
Further evidence for this rich chain of conjectured dualities and a discussion of  branes and holography in these theories can be  found in \cite{Hull:1998fh,Hull:1999mt,Dijkgraaf:2016lym}.

\subsection{Further Supergravities in $d<9$}

Any of these theories in 10 or 11 dimensions with signature $S+T$ can be further reduced on $p$ spatial dimensions and $q$ time dimensions
with $p\leq S$ and $q\leq T$ to give a theory in  $s+t$ dimensions with $s=S-p$ and $t=T-q$. This gives variants of the supergravities considered in the earlier sections in different spacetime signatures. In each case the duality group $G$ is the same as the ones considered earlier but with the subgroup $H$ in the coset $G/H$ replaced by a different real form of the same group to give a coset of the same dimension, but possibly a different signature. For example, for reduction from 11 to 7 dimensions, $G=SL(5,\mathbb{R})$ and the different real forms of $H$ that arise are
$H=SO(r,5-r)$ for $r=0,1,2$. 
 For the standard reduction in section \ref{dualsugra}, $H=SO(5)$, for the timelike reduction in section \ref{timelike} it is $H=SO(4,1)$ while for the reduction of the 11* theory on $T^{3,1}$ to give a supergravity in $6+1$ dimensions  it  is $H=SO(3,2)$. For $G=E_6$ the possible forms of $H$ are $U S p (8), U {Sp} (4, 4), S p (8, \mathbb{R})$, for $G=E_7$ the possible forms of $H$ are
$S U (8),S U (4, 4), S L (8, \mathbb{R}), S U^{\ast} (8)$
while for $G=E_8$ the possible forms of $H$ are $S p i n (16),S p i n (8, 8), S p i n^{\ast} (16)$.


\begin{thebibliography}{99}
\bibitem{Cremmer:1978ds}
E.~Cremmer and B.~Julia,
``The N=8 Supergravity Theory. 1. The Lagrangian,''
Phys. Lett. B \textbf{80} (1978), 48

\bibitem{Cremmer:1979up}
E.~Cremmer and B.~Julia,
``The SO(8) Supergravity,''
Nucl. Phys. B \textbf{159} (1979), 141-212

\bibitem{Cremmer:1980gs}
E.~Cremmer,
``Supergravities in 5 Dimensions,''
in {\it Superspace and Supergravity}, Eds. S.W. Hawking and M. Rocek (Cambridge Univ. Press,
1981).

\bibitem{Julia:1980gr}
B.~Julia,
``GROUP DISINTEGRATIONS,''
in {\it Superspace and Supergravity}, Eds. S.W. Hawking and M. Rocek (Cambridge Univ. Press,
1981).



\bibitem{Julia:1981wc}
B.~Julia,
``Infinite lie algebras in physics,'' 
talk at the Johns Hopkins Workshop on Particle theory
1981,
LPTENS-81-14.



\bibitem{Cremmer:1997ct}
E.~Cremmer, B.~Julia, H.~Lu and C.~N.~Pope,
``Dualization of dualities. 1,''
Nucl. Phys. B \textbf{523} (1998), 73-144
[arXiv:hep-th/9710119 [hep-th]].


\bibitem{Hull:1994ys}
C.~M.~Hull and P.~K.~Townsend,
``Unity of superstring dualities,''
Nucl. Phys. B \textbf{438} (1995), 109-137
[arXiv:hep-th/9410167 [hep-th]].



\bibitem{Samtleben:2023ivs}
H.~Samtleben,
``11D Supergravity and Hidden Symmetries,''
[arXiv:2303.12682 [hep-th]].


\bibitem{Damour:2002cu}
T.~Damour, M.~Henneaux and H.~Nicolai,
``E(10) and a `small tension expansion' of M theory,''
Phys. Rev. Lett. \textbf{89} (2002), 221601
[arXiv:hep-th/0207267 [hep-th]].

\bibitem{West:2003fc}
P.~C.~West,
``E(11), SL(32) and central charges,''
Phys. Lett. B \textbf{575} (2003), 333-342
[arXiv:hep-th/0307098 [hep-th]];
P.~C.~West,
``E(11) origin of brane charges and U-duality multiplets,''
JHEP \textbf{08} (2004), 052
[arXiv:hep-th/0406150 [hep-th]];
A.~G.~Tumanov and P.~West,
``E11 in 11D,''
Phys. Lett.  B  \textbf{758} (2016), 278-285
[arXiv:1601.03974 [hep-th]].


\bibitem{Hull:1985pq}
C.~M.~Hull, A.~Karlhede, U.~Lindstrom and M.~Rocek,
``Nonlinear $\sigma$ Models and Their Gauging in and Out of Superspace,''
Nucl. Phys. B \textbf{266} (1986), 1-44

\bibitem{Gomez-Fayren:2024cpl}
C.~G\'omez-Fayr\'en, T.~Ort\'\i{}n and M.~Zatti,
``Gravitational higher-form symmetries and the origin of hidden symmetries in Kaluza-Klein compactifications,''
SciPost Phys. Core \textbf{8} (2025), 010
[arXiv:2405.16706 [hep-th]].

\bibitem{Tanii:1984zk}
Y.~Tanii,
``$N=8$ Supergravity in Six-dimensions,''
Phys. Lett. B \textbf{145} (1984), 197-200

\bibitem{Gaillard:1981rj}
M.~K.~Gaillard and B.~Zumino,
``Duality Rotations for Interacting Fields,''
Nucl. Phys. B \textbf{193} (1981), 221-244

\bibitem{Cremmer:1978km}
E.~Cremmer, B.~Julia and J.~Scherk,
``Supergravity Theory in Eleven-Dimensions,''
Phys. Lett. B \textbf{76} (1978), 409-412

\bibitem{Maharana:1992my}
J.~Maharana and J.~H.~Schwarz,
``Noncompact symmetries in string theory,''
Nucl. Phys. B \textbf{390} (1993), 3-32
[arXiv:hep-th/9207016 [hep-th]].

\bibitem{Ehlers:1959aug}
J.~Ehlers,
``Transformations of static exterior solutions of Einstein's gravitational field equations into different solutions by means of conformal mapping,''
Colloq. Int. CNRS \textbf{91} (1962), 275-284

\bibitem{Geroch:1972yt}
R.~P.~Geroch,
``A Method for generating new solutions of Einstein's equation. 2,''
J. Math. Phys. \textbf{13} (1972), 394-404

\bibitem{Cremmer:1998px}
E.~Cremmer, B.~Julia, H.~Lu and C.~N.~Pope,
``Dualization of dualities. 2. Twisted self-duality of doubled fields, and superdualities,''
Nucl. Phys. B \textbf{535} (1998), 242-292
[arXiv:hep-th/9806106 [hep-th]].


\bibitem{Duff:1983vj}
M.~J.~Duff, B.~E.~W.~Nilsson and C.~N.~Pope,
``Compactification of $d=11$ Supergravity on $K$(3) X U(3),''
Phys. Lett. B \textbf{129} (1983), 39


\bibitem{Mizoguchi:1998wv}
S.~Mizoguchi and N.~Ohta,
``More on the similarity between D = 5 simple supergravity and M theory,''
Phys. Lett. B \textbf{441} (1998), 123-132
[arXiv:hep-th/9807111 [hep-th]].


\bibitem{Cremmer:1999du}
E.~Cremmer, B.~Julia, H.~Lu and C.~N.~Pope,
``Higher dimensional origin of D = 3 coset symmetries,''
[arXiv:hep-th/9909099 [hep-th]].

\bibitem{Geroch:1970nt}
R.~P.~Geroch,
``A Method for generating solutions of Einstein's equations,''
J. Math. Phys. \textbf{12} (1971), 918-924


\bibitem{Hull:1998br}
C.~M.~Hull and B.~Julia,
``Duality and moduli spaces for timelike reductions,''
Nucl. Phys. B \textbf{534} (1998), 250-260
[arXiv:hep-th/9803239 [hep-th]].



\bibitem{Gibbons:1979xm}
G.~W.~Gibbons and S.~W.~Hawking,
``Classification of Gravitational Instanton Symmetries,''
Commun. Math. Phys. \textbf{66} (1979), 291-310

\bibitem{Breitenlohner:1987dg}
P.~Breitenlohner, D.~Maison and G.~W.~Gibbons,
``Four-Dimensional Black Holes from Kaluza-Klein Theories,''
Commun. Math. Phys. \textbf{120} (1988), 295



\bibitem{HullTalk}
C.~M.~Hull, ``What was string theory? The story of the string duality revolution," talk given in KITP Program: {\it What is String Theory? Weaving Perspectives Together},
https://online.kitp.ucsb.edu/online/strings24/hull/

\bibitem{Obers:1998fb}
N.~A.~Obers and B.~Pioline,
``U duality and M theory,''
Phys. Rept. \textbf{318} (1999), 113-225
[arXiv:hep-th/9809039 [hep-th]].

\bibitem{Sen:1993zi}
A.~Sen,
``Magnetic monopoles, Bogomolny bound and SL(2,Z) invariance in string theory,''
Mod. Phys. Lett. A \textbf{8} (1993), 2023-2036
[arXiv:hep-th/9303057 [hep-th]].

\bibitem{Sen:1993sx}
A.~Sen,
``SL(2,Z) duality and magnetically charged strings,''
Int. J. Mod. Phys. A \textbf{8} (1993), 5079-5094
[arXiv:hep-th/9302038 [hep-th]].


\bibitem{Sen:1992fr}
A.~Sen,
``Electric magnetic duality in string theory,''
Nucl. Phys. B \textbf{404} (1993), 109-126
[arXiv:hep-th/9207053 [hep-th]].


\bibitem{Giveon:1994fu}
A.~Giveon, M.~Porrati and E.~Rabinovici,
``Target space duality in string theory,''
Phys. Rept. \textbf{244} (1994), 77-202
[arXiv:hep-th/9401139 [hep-th]].

\bibitem{Narain:1985jj}
K.~S.~Narain,
``New Heterotic String Theories in Uncompactified Dimensions \ensuremath{<} 10,''
Phys. Lett. B \textbf{169} (1986), 41-46


\bibitem{Narain:1986am}
K.~S.~Narain, M.~H.~Sarmadi and E.~Witten,
``A Note on Toroidal Compactification of Heterotic String Theory,''
Nucl. Phys. B \textbf{279} (1987), 369-379


\bibitem{Witten:1978mh}
E.~Witten and D.~I.~Olive,
``Supersymmetry Algebras That Include Topological Charges,''
Phys. Lett. B \textbf{78} (1978), 97-101

\bibitem{Montonen:1977sn}
C.~Montonen and D.~I.~Olive,
``Magnetic Monopoles as Gauge Particles?,''
Phys. Lett. B \textbf{72} (1977), 117-120

\bibitem{Osborn:1979tq}
H.~Osborn,
``Topological Charges for N=4 Supersymmetric Gauge Theories and Monopoles of Spin 1,''
Phys. Lett. B \textbf{83} (1979), 321-326


\bibitem{Sen:1994yi}
A.~Sen,
``Dyon - monopole bound states, selfdual harmonic forms on the multi - monopole moduli space, and SL(2,Z) invariance in string theory,''
Phys. Lett. B \textbf{329} (1994), 217-221
[arXiv:hep-th/9402032 [hep-th]].


\bibitem{Font:1990gx}
A.~Font, L.~E.~Ibanez, D.~Lust and F.~Quevedo,
``Strong - weak coupling duality and nonperturbative effects in string theory,''
Phys. Lett. B \textbf{249} (1990), 35-43

\bibitem{Schwarz:1993mg}
J.~H.~Schwarz and A.~Sen,
``Duality symmetries of 4-D heterotic strings,''
Phys. Lett. B \textbf{312} (1993), 105-114
[arXiv:hep-th/9305185 [hep-th]].

\bibitem{Schwarz:1993vs}
J.~H.~Schwarz and A.~Sen,
``Duality symmetric actions,''
Nucl. Phys. B \textbf{411} (1994), 35-63
[arXiv:hep-th/9304154 [hep-th]].

\bibitem{Gibbons:1981ux}
G.~W.~Gibbons,
``Soliton States and Central Charges in Extended Supergravity Theories,''
Lect. Notes Phys. \textbf{160} (1982), 145-151

\bibitem{Gibbons:1982fy}
G.~W.~Gibbons and C.~M.~Hull,
``A Bogomolny Bound for General Relativity and Solitons in N=2 Supergravity,''
Phys. Lett. B \textbf{109} (1982), 190-194

\bibitem{Dai:1989ua}
J.~Dai, R.~G.~Leigh and J.~Polchinski,
``New Connections Between String Theories,''
Mod. Phys. Lett. A \textbf{4} (1989), 2073-2083





\bibitem{Townsend:1995kk}
P.~K.~Townsend,
``The eleven-dimensional supermembrane revisited,''
Phys. Lett. B \textbf{350} (1995), 184-187
[arXiv:hep-th/9501068 [hep-th]].


\bibitem{Witten:1995ex}
E.~Witten,
``String theory dynamics in various dimensions,''
Nucl. Phys. B \textbf{443} (1995), 85-126
[arXiv:hep-th/9503124 [hep-th]].

\bibitem{Polchinski:1995mt}
J.~Polchinski,
``Dirichlet Branes and Ramond-Ramond charges,''
Phys. Rev. Lett. \textbf{75} (1995), 4724-4727
[arXiv:hep-th/9510017 [hep-th]].

\bibitem{Hull:1998vg}
C.~M.~Hull,
``Timelike T duality, de Sitter space, large N gauge theories and topological field theory,''
JHEP \textbf{07} (1998), 021
[arXiv:hep-th/9806146]

\bibitem{Hull:1998ym}
C.~M.~Hull,
``Duality and the signature of space-time,''
JHEP \textbf{11} (1998), 017
[arXiv:hep-th/9807127]

\bibitem{Kugo:1982bn}
T.~Kugo and P.~K.~Townsend,
``Supersymmetry and the Division Algebras,''
Nucl. Phys. B \textbf{221} (1983), 357-380

\bibitem{Hull:1998fh}
C.~M.~Hull and R.~R.~Khuri,
``Branes, times and dualities,''
Nucl. Phys. B \textbf{536} (1998), 219-244
[arXiv:hep-th/9808069 [hep-th]].

\bibitem{Hull:1999mt}
C.~M.~Hull and R.~R.~Khuri,
``World volume theories, holography, duality and time,''
Nucl. Phys. B \textbf{575} (2000), 231-254
[arXiv:hep-th/9911082 [hep-th]].

\bibitem{Dijkgraaf:2016lym}
R.~Dijkgraaf, B.~Heidenreich, P.~Jefferson and C.~Vafa,
``Negative Branes, Supergroups and the Signature of Spacetime,''
JHEP \textbf{02} (2018), 050
[arXiv:1603.05665 [hep-th]].

\bibitem{Duff:2006ix}
M.~J.~Duff and J.~Kalkkinen,
``Signature reversal invariance,''
Nucl. Phys. B \textbf{758} (2006), 161-184
[arXiv:hep-th/0605273 [hep-th]].


\end{thebibliography}
\end{document}